\documentclass[twocolumn]{IEEEtran} 

\def\BibTeX{{\rm B\kern-.05em{\sc i\kern-.025em b}\kern-.08em
    T\kern-.1667em\lower.7ex\hbox{E}\kern-.125emX}}

\begin{document}

\title{Towards
 Electronic Shopping
 of Composite Product}

\author{Mark Sh. Levin
\thanks{Mark Sh. Levin:
 http://www.mslevin.iitp.ru
  (e-mail: mslevin@acm.org).}
}


\maketitle

\begin{abstract}
 In the paper,
 frameworks for electronic shopping of composite
 (modular) products are described:
 (a) multicriteria selection (product is considered as a whole
 system, it is a traditional approach),
 (b) combinatorial synthesis (composition)
 of the product from its components,
 (c) aggregation of the product from several selected
 products/prototypes.
%
 The following product model is examined:
%
 (i) general tree-like structure,
 (ii) set of system parts/components (leaf nodes),
 (iii) design alternatives (DAs) for each component,
 (iv) ordinal priorities for DAs, and
 (v) estimates of compatibility  between DAs for different
 components.
%
 The combinatorial synthesis is realized
 as morphological design of a composite (modular) product or an
 extended composite product (e.g., product and support services as financial instruments).
 Here the solving process is based on Hierarchical Morphological Multicriteria Design
 (HMMD):
 (i) multicriteria selection of alternatives for system
 parts,
 (ii) composing the selected alternatives
 into a resultant combination
 (while taking into
 account ordinal quality of the alternatives above
 and their compatibility).
%
 The aggregation framework is based on
 consideration of
  aggregation procedures, for example:
 (i) addition procedure: design of a products substructure
 or an extended substructure (``kernel'') and addition of
  elements, and
%
%
 (ii) design procedure:
 design of the composite solution based on
 all elements of product superstructure.
%
%
 Applied numerical examples
 (e.g., composite product, extended composite product, product repair plan,
 and product trajectory)
 illustrate the proposed approaches.
\end{abstract}

\begin{keywords}
 Electronic shopping,
 modular products,
 morphological design,
 combinatorial optimization,
 multicriteria decision making,
 aggregation,
 customer centric design
\end{keywords}

\newcounter{cms}
\setlength{\unitlength}{1mm}


\section{Introduction}

 In recent decade, the significance of electronic shopping
 and usage of corresponding recommender systems
 is
 increased (e.g.,
 \cite{haubl00},
 \cite{kolakota96},
 \cite{lu07},
 \cite{ricci11},
 \cite{rust07},
 \cite{standing00},
 \cite{tseng07}).
 Here it is reasonable to point out the following basic directions:
%
%
  {\it 1.} various recommender systems
 (e.g., \cite{ado05},
 \cite{balab97},
 \cite{cho02},
 \cite{lee02},
 \cite{perug04},
 \cite{resnick97},
 \cite{ricci11},
 \cite{wang04});
  {\it 2.} electronic services for business in electronic environments
  (e.g.,
 \cite{alonso04},
 \cite{bat01},
 \cite{castro06},
 \cite{leymann02},
   \cite{orman07},
 \cite{rust07},
 \cite{standing00},
 \cite{yolum05});
  {\it 3.} issues of distributed information retrieval and integration
 (e.g.,
 \cite{kno01},
 \cite{roth02},
 \cite{ste00});
%
 {\it 4.} multistage information retrieval
 (e.g.,
  \cite{pit02}, \cite{yu02});
  {\it 5.} design of websites for electronic shopping
 (e.g.,
 \cite{ceri00},
 \cite{tseng07}, \cite{youwei06});
 {\it 6.} usage of ontology approaches to web services
 (e.g., \cite{roman05});
 {\it 7.} adaptation of Web sites and systems
  (e.g., \cite{perk00});
%
 {\it 8.} personalization of Web-based systems
 (search and recommender systems, etc.)
 (e.g.,
 \cite{ado05a},
 \cite{cho02},
 \cite{eir03},
 \cite{pit02},
 \cite{wang04});
 {\it 9.} usage of operations research methods and/or AI techniques
 (e.g., \cite{geoff01},
 \cite{sodhi01});
 and
 {\it 10.} some efforts in Web-based product design,
 e.g.,
   Web-based combining a composite product
   (\cite{mad06}, \cite{sid05}),
    special designer-buyer-supplier interfaces over the Web to facilitate
    product development
    (e.g., \cite{dowlatshahi97}, \cite{huang00}).
%
 A simplified scheme 'user-electronic resources' is
 presented in Fig. 1.
%
%
 Note, the development of contemporary Web-based systems
 is targeted to and based on
 Web-based support systems (e.g., \cite{lu07}).
%
 Decision support tools may be used at different levels:
 (i) interface, (ii) search engines, and (iii) data bases.

 In our opinion, some basic problems in electronic shopping are
 the following (Table 1):
(i) searching for a product on the basis of requirements
 (criteria) or user preferences,
 (ii) selection of a product on the basis
  of multicriteria decision making, and
 (iii) selection of product(s) under some constraints
 (e.g., multicriteria knapsack problem),
 (iv) multiple selection in several databases
 under a total resource constraint(s)
 (multiple choice knapsack problem),
 (v) design of a configuration for a modular product
 (e.g., morphological composition of the product from its components),
 and
 (vi) aggregation of  selected modular solutions
 (as consensus, median-like solution).

%
%


\begin{center}
\begin{picture}(80,55)
\put(18.5,00){\makebox(0,0)[bl]{Fig. 1. General
 framework}}


\put(4,37){\circle{4}}

\put(4,25){\line(0,1){10}}

\put(0,31){\line(1,0){8}}

\put(4,25){\line(-1,-1){4}}

\put(4,25){\line(1,-1){4}}

\put(10,05){\line(1,0){10}} \put(10,54){\line(1,0){10}}
\put(10,05){\line(0,1){49}} \put(20,05){\line(0,1){49}}

\put(11,33){\makebox(0,0)[bl]{Inter-}}
\put(11,29){\makebox(0,0)[bl]{face}}

\put(25,05){\line(1,0){55}} \put(25,16){\line(1,0){55}}
\put(25,05){\line(0,1){11}} \put(80,05){\line(0,1){11}}

\put(25.5,05){\line(0,1){11}} \put(79.5,05){\line(0,1){11}}

\put(30,12){\makebox(0,0)[bl]{Support tools (search/retrieval, }}
\put(30,09){\makebox(0,0)[bl]{ranking/selection, composition/}}
\put(30,06){\makebox(0,0)[bl]{synthesis, aggregation)}}

\put(25,10.5){\vector(-1,0){5}}

\put(35,16){\vector(0,1){4}}

\put(56.5,16){\vector(0,1){6}}


\put(66,16){\line(0,1){26}} \put(66,42){\vector(1,1){4}}
\put(66,22){\vector(1,1){4}}

\put(25,20){\line(1,0){20}}\put(25,54){\line(1,0){20}}
\put(25,20){\line(0,1){34}} \put(45,20){\line(0,1){34}}

\put(25.5,20.5){\line(1,0){19}}\put(25.5,53.5){\line(1,0){19}}
\put(25.5,20.5){\line(0,1){33}} \put(44.5,20.5){\line(0,1){33}}

\put(26,49){\makebox(0,0)[bl]{Support}}
\put(26,45){\makebox(0,0)[bl]{recommen-}}
\put(26,41){\makebox(0,0)[bl]{dation}}
\put(26,36.7){\makebox(0,0)[bl]{problems}}
\put(26,33){\makebox(0,0)[bl]{(selection,}}
\put(26,29){\makebox(0,0)[bl]{design/}}
\put(26,25){\makebox(0,0)[bl]{composition,}}
\put(26,21.3){\makebox(0,0)[bl]{aggregation)}}

\put(20,37){\line(1,0){5}}


\put(56.5,37){\oval(13,30)} \put(56.5,37){\oval(12,29)}

\put(51,39){\makebox(0,0)[bl]{Search}}
\put(51,35){\makebox(0,0)[bl]{engines}}

\put(45,37){\line(1,0){5}}

\put(70,22){\line(1,0){10}}\put(70,32){\line(1,0){10}}
\put(70,22){\line(0,1){10}} \put(80,22){\line(0,1){10}}
\put(71,28){\makebox(0,0)[bl]{Data}}
\put(71,24){\makebox(0,0)[bl]{base}}

\put(63,27){\line(1,0){7}}

\put(72.5,36.6){\makebox(0,0)[bl]{{\bf . . .}}}

\put(70,42){\line(1,0){10}}\put(70,52){\line(1,0){10}}
\put(70,42){\line(0,1){10}} \put(80,42){\line(0,1){10}}
\put(71,48){\makebox(0,0)[bl]{Data}}
\put(71,44){\makebox(0,0)[bl]{base}}

\put(63,47){\line(1,0){7}}

\end{picture}
\end{center}

\begin{center}
\begin{picture}(80,86)

\put(15,81){\makebox(0,0)[bl] {Table 1. Problems and methods}}

\put(0,1){\line(1,0){80}} \put(0,73){\line(1,0){80}}
\put(0,79){\line(1,0){80}}

\put(0,1){\line(0,1){78}} \put(41,1){\line(0,1){78}}
\put(80,1){\line(0,1){78}}

\put(01,74.7){\makebox(0,0)[bl]{Problems}}

\put(42,74.7){\makebox(0,0)[bl]{Models/methods }}

\put(1,68){\makebox(0,0)[bl]{1.Searching for a product}}
\put(42,68.5){\makebox(0,0)[bl]{Information retrieval}}

\put(1,63){\makebox(0,0)[bl]{2.Multicriteria selection}}
\put(1,59){\makebox(0,0)[bl]{of a product}}

\put(42,63){\makebox(0,0)[bl]{Multicriteria ranking}}

\put(1,54){\makebox(0,0)[bl]{3.Selection of products}}
\put(1,50.4){\makebox(0,0)[bl]{under resource}}
\put(1,46){\makebox(0,0)[bl]{constraint(s)}}

\put(42,54){\makebox(0,0)[bl]{Knapsack-like problems}}

\put(1,41){\makebox(0,0)[bl]{4.Multi-selection of}}
\put(1,36.7){\makebox(0,0)[bl]{several products under}}
\put(1,33){\makebox(0,0)[bl]{resource constraint(s)}}

\put(42,41){\makebox(0,0)[bl]{Multiple choice problem}}
\put(42,37){\makebox(0,0)[bl]{(including multicriteria}}
\put(42,33){\makebox(0,0)[bl]{multiple choice problem)}}


\put(1,28){\makebox(0,0)[bl]{5.Design of configuration}}
\put(1,24){\makebox(0,0)[bl]{for composite (modular)}}
\put(1,20){\makebox(0,0)[bl]{product, extended product}}

\put(42,28){\makebox(0,0)[bl]{Morphological design,}}
\put(42,24){\makebox(0,0)[bl]{multiple choice problem,}}
\put(42,20){\makebox(0,0)[bl]{AI techniques, etc.}}

\put(1,15){\makebox(0,0)[bl]{6.Aggregation of several }}

\put(1,11){\makebox(0,0)[bl]{selected products}}


\put(42,15){\makebox(0,0)[bl]{Aggregation methods}}

\put(42,11){\makebox(0,0)[bl]{(e.g., consensus,}}

\put(42,7){\makebox(0,0)[bl]{median structure,}}


\put(42,3){\makebox(0,0)[bl]{new design)}}


\end{picture}
\end{center}

 This paper
  describes
 three basic frameworks for electronic shopping of composite (modular) products:
 {\it 1.} multicriteria selection (product is considered as a whole
 system, it is a traditional approach);
%
 {\it 2.} combinatorial synthesis (composition)
 of the product from its components
 (i.e., design/synthesis of
 configuration for the
 modular product and extended modular product); and
 {\it 3.} design of an aggregated product on the basis
   of several selected products/prototypes.

 The following model of the composite (modular) product
 is examined:
 (\cite{lev98}, \cite{lev05},
 \cite{lev06}, \cite{lev11agg}):
 (i) tree-like system structure,
 (ii) set of leaf nodes as system parts/components,
 (iii) design alternatives (DAs) for each system part/component,
 (iv) ordinal priorities for DAs, and
 (v) estimates of compatibility  between DAs for different
 system parts/components.

 Our combinatorial synthesis is
 based on morphological design of the composite (modular) product or
 extended composite product (e.g., product and support services as financial instruments).
 Here  Hierarchical Morphological Multicriteria Design
 (HMMD) approach is used
 (\cite{lev98}, \cite{lev05}, \cite{lev06}):
 (i) multicriteria selection of alternatives for system parts,
 (ii) composing the selected alternatives into a resultant combination
 (while taking into account ordinal quality of the alternatives above
 and their compatibility).
%

%
 In this paper,
 two aggregation procedures are considered \cite{lev11agg}:
%
 (i) addition (extension) procedure:
 design of a products substructure or an extended substructure
 (``kernel'') and addition of elements, and
%
%
 (ii) design procedure:
 design of the composite solution based on
 all elements of product superstructure.


 Applied numerical examples
 (composite products, extended composite product, product repair plan,
 product trajectory)
 illustrate the proposed approaches.

 Note similar type of e-commerce
 is considered as "designing while shopping" \cite{lu06}.
 Generally, our combinatorial
 synthesis approaches is based on
 three basic types of combinatorial solving schemes (Table 1):

 (1) multiple choice knapsack problem
 (\cite{gar79}, \cite{kellerer04}, \cite{mar90}),

  (2) Hierarchical Morphological Multicriteria Design
 (HMMD) approach
 (e.g., \cite{lev98},
 \cite{lev05},
  \cite{lev06}),
 and

 (3) aggregation procedures \cite{lev11agg}.

 The combinatorial approaches can be considered as a fundamental
 for
 two processes:
 (a) product design (i.e., synthesis, composition, aggregation)
 and
 (b) accumulation and representation of
 customers requirements, preferences, and needs.
%
%

 A preliminary material of the paper was published as conference paper \cite{lev08a},
 a simplified example of product aggregation was presented in \cite{lev11agg}.

\section{Structured Model of Product}

 The following hierarchical multi-layer model
  ``morphological tree''
 for composite product is examined
 (\cite{lev98}, \cite{lev05}, \cite{lev06}, \cite{lev11agg})
 (Fig. 2):

 (i) tree-like system model ({\bf T}),

 (ii) set of leaf nodes as basic system parts/components
 (e.g., \(\{P_{1},...,P_{i},...,P_{m}\}\)),

 (iii) sets of design alternatives (DAs) for each leaf node,

 (iv) rankings of DAs (i.e., ordinal priorities)
 ({\bf R}), and

 (v) compatibility estimates between DAs for different leaf nodes
 ({\bf I}).

 This ``morphological tree'' model
 is a version of ``and-or tree''.

\begin{center}
\begin{picture}(76,63)
\put(06,00){\makebox(0,0)[bl]{Fig. 2. Architecture of modular
 product
 \cite{lev11agg} }}


\put(38,61){\circle*{3}}


\put(23,49.5){\makebox(0,0)[bl]{Hierarchy of product}}
\put(24,46){\makebox(0,0)[bl]{(tree-like structure)
 }}

\put(02,39){\makebox(0,0)[bl]{System parts/}}
\put(02,36){\makebox(0,0)[bl]{components}}
\put(02,33){\makebox(0,0)[bl]{(leaf nodes)
 }}

\put(04,33){\line(0,-1){5}}

\put(05,33){\line(2,-3){5.4}}

\put(6,33){\line(3,-1){24}}


\put(27,39){\makebox(0,0)[bl]{Sets of design}}
\put(27,36.5){\makebox(0,0)[bl]{alternatives}}

\put(33,35.6){\line(-1,-1){16}} \put(43,35.6){\line(1,-1){16}}


\put(54,39){\makebox(0,0)[bl]{Compatibility}}
\put(54,36){\makebox(0,0)[bl]{among}}
\put(54,33){\makebox(0,0)[bl]{design}}
\put(54,30.5){\makebox(0,0)[bl]{alternatives
 }}

\put(62,30){\line(-1,-1){21}}


\put(00,23){\line(1,0){76}}

\put(00,42){\line(2,1){38}} \put(76,42){\line(-2,1){38}}

\put(00,23){\line(0,1){19}} \put(76,23){\line(0,1){19}}

\put(01,24.5){\makebox(0,0)[bl]{\(P_{1}\)}}

\put(03,23){\circle*{2}} \put(03,18){\oval(6,8)}

\put(12,24.5){\makebox(0,0)[bl]{\(P_{2}\)}}

\put(13,23){\circle*{2}} \put(13,18){\oval(6,8)}

\put(30.5,24){\makebox(0,0)[bl]{\(P_{i-1}\)}}

\put(33,23){\circle*{2}} \put(33,18){\oval(6,8)}

\put(42.5,24.5){\makebox(0,0)[bl]{\(P_{i}\)}}

\put(43,23){\circle*{2}} \put(43,18){\oval(6,8)}

\put(59.5,24.5){\makebox(0,0)[bl]{\(P_{m-1}\)}}

\put(63,23){\circle*{2}} \put(63,18){\oval(6,8)}

\put(70.6,24.5){\makebox(0,0)[bl]{\(P_{m}\)}}

\put(73,23){\circle*{2}} \put(73,18){\oval(6,8)}

\put(20,17.5){\makebox(0,0)[bl]{. . .}}
\put(50,17.5){\makebox(0,0)[bl]{. . .}}

\put(20,9){\makebox(0,0)[bl]{. . .}}
\put(50,9){\makebox(0,0)[bl]{. . .}}
\put(06,06){\line(1,0){04}} \put(06,13){\line(1,0){04}}
\put(06,06){\line(0,1){7}} \put(10,06){\line(0,1){7}}

\put(07,06){\line(0,1){7}} \put(08,06){\line(0,1){7}}
\put(09,06){\line(0,1){7}}

\put(36,06){\line(1,0){04}} \put(36,13){\line(1,0){04}}
\put(36,06){\line(0,1){7}} \put(40,06){\line(0,1){7}}

\put(37,06){\line(0,1){7}} \put(38,06){\line(0,1){7}}
\put(39,06){\line(0,1){7}}

\put(66,06){\line(1,0){04}} \put(66,13){\line(1,0){04}}
\put(66,06){\line(0,1){7}} \put(70,06){\line(0,1){7}}

\put(67,06){\line(0,1){7}} \put(68,06){\line(0,1){7}}
\put(69,06){\line(0,1){7}}


\end{picture}
\end{center}

 Further, two simplified illustrative examples of structured models are
 presented (estimates have only illustrative character).
 Fig. 3 depicts a three-part motor vehicle
 (some priorities of DAs are depicted in  parentheses,
 \(1\) corresponds to the best level):
 ~{\it 1.} body \(A\)
 (sedan \(A_{1}\),
 universal \(A_{2}\),
 jeep \(A_{3}\),
 pickup \(A_{4}\), and
 sport \(A_{5}\));
 ~{\it 2.} engine \(B\)
 (diesel \(B_{1}\),
 gasoline \(B_{2}\),
 electric \(B_{3}\), and
 hydrogenous \(B_{4}\)); and
 ~{\it 3.} equipment \(C\)
 (basic alternative \(C_{1}\),
 computer control \(C_{2}\), and
 computer control \& GPS-linked \(C_{3}\)).
 Table 2 contains ordinal estimates of compatibility between DAs
 for different product components which are based on expert
 judgment
 (\(3\) corresponds to the best level of compatibility,
 \(0\) corresponds to incompatibility).

\begin{center}
\begin{picture}(40,40)
\put(03,0){\makebox(0,0)[bl] {Fig. 3. Motor vehicle}}



\put(0,4){\makebox(0,8)[bl]{\(A_{5}(2)\)}}
\put(0,08){\makebox(0,8)[bl]{\(A_{4}(3)\)}}
\put(0,12){\makebox(0,8)[bl]{\(A_{3}(2)\)}}
\put(0,16){\makebox(0,8)[bl]{\(A_{2}(3)\)}}
\put(0,20){\makebox(0,8)[bl]{\(A_{1}(1)\)}}

\put(13,08){\makebox(0,8)[bl]{\(B_{4}(3)\)}}
\put(13,12){\makebox(0,8)[bl]{\(B_{3}(2)\)}}
\put(13,16){\makebox(0,8)[bl]{\(B_{2}(1)\)}}
\put(13,20){\makebox(0,8)[bl]{\(B_{1}(1)\)}}

\put(27,12){\makebox(0,8)[bl]{\(C_{3}(3)\)}}
\put(27,16){\makebox(0,8)[bl]{\(C_{2}(2)\)}}
\put(27,20){\makebox(0,8)[bl]{\(C_{1}(1)\)}}

\put(1,25){\circle*{1.6}} \put(16,25){\circle*{1.6}}
\put(31,25){\circle*{1.6}}

\put(1,30){\line(0,-1){04}} \put(16,30){\line(0,-1){04}}
\put(31,30){\line(0,-1){04}}

\put(1,30){\line(1,0){30}}

\put(07,30){\line(0,1){6}} \put(07,36){\circle*{3}}

\put(02,27){\makebox(0,8)[bl]{\(A\) }}
\put(12,27){\makebox(0,8)[bl]{\(B\) }}
\put(27,27){\makebox(0,8)[bl]{\(C\) }}

\put(11,36){\makebox(0,8)[bl]{\(S = A \star B \star C \) }}

\put(09,32){\makebox(0,8)[bl]{\(S_{1}=A_{1}\star B_{3}\star
C_{2}\)}}

\end{picture}
%
\begin{picture}(42,50)

\put(4,46){\makebox(0,0)[bl]{Table 2. Compatibility}}

\put(00,0){\line(1,0){42}} \put(00,38){\line(1,0){42}}
\put(00,44){\line(1,0){42}}

\put(00,0){\line(0,1){44}} \put(07,0){\line(0,1){44}}
\put(42,0){\line(0,1){44}}

\put(01,34){\makebox(0,0)[bl]{\(A_{1}\)}}
\put(01,30){\makebox(0,0)[bl]{\(A_{2}\)}}
\put(01,26){\makebox(0,0)[bl]{\(A_{3}\)}}
\put(01,22){\makebox(0,0)[bl]{\(A_{4}\)}}
\put(01,18){\makebox(0,0)[bl]{\(A_{5}\)}}

\put(01,14){\makebox(0,0)[bl]{\(B_{1}\)}}
\put(01,10){\makebox(0,0)[bl]{\(B_{2}\)}}
\put(01,06){\makebox(0,0)[bl]{\(B_{3}\)}}
\put(01,02){\makebox(0,0)[bl]{\(B_{4}\)}}

\put(12,38){\line(0,1){6}} \put(17,38){\line(0,1){6}}
\put(22,38){\line(0,1){6}} \put(27,38){\line(0,1){6}}
\put(32,38){\line(0,1){6}} \put(37,38){\line(0,1){6}}

\put(07.4,40){\makebox(0,0)[bl]{\(B_{1}\)}}
\put(12.4,40){\makebox(0,0)[bl]{\(B_{2}\)}}
\put(17.4,40){\makebox(0,0)[bl]{\(B_{3}\)}}
\put(22.4,40){\makebox(0,0)[bl]{\(B_{4}\)}}
\put(27.4,40){\makebox(0,0)[bl]{\(C_{1}\)}}
\put(32.4,40){\makebox(0,0)[bl]{\(C_{2}\)}}
\put(37.4,40){\makebox(0,0)[bl]{\(C_{3}\)}}

\put(09,34){\makebox(0,0)[bl]{\(3\)}}
\put(14,34){\makebox(0,0)[bl]{\(3\)}}
\put(19,34){\makebox(0,0)[bl]{\(2\)}}
\put(24,34){\makebox(0,0)[bl]{\(1\)}}
\put(29,34){\makebox(0,0)[bl]{\(2\)}}
\put(34,34){\makebox(0,0)[bl]{\(3\)}}
\put(39,34){\makebox(0,0)[bl]{\(2\)}}

\put(09,30){\makebox(0,0)[bl]{\(3\)}}
\put(14,30){\makebox(0,0)[bl]{\(2\)}}
\put(19,30){\makebox(0,0)[bl]{\(2\)}}
\put(24,30){\makebox(0,0)[bl]{\(2\)}}
\put(29,30){\makebox(0,0)[bl]{\(1\)}}
\put(34,30){\makebox(0,0)[bl]{\(2\)}}
\put(39,30){\makebox(0,0)[bl]{\(3\)}}

\put(09,26){\makebox(0,0)[bl]{\(3\)}}
\put(14,26){\makebox(0,0)[bl]{\(3\)}}
\put(19,26){\makebox(0,0)[bl]{\(0\)}}
\put(24,26){\makebox(0,0)[bl]{\(0\)}}
\put(29,26){\makebox(0,0)[bl]{\(1\)}}
\put(34,26){\makebox(0,0)[bl]{\(3\)}}
\put(39,26){\makebox(0,0)[bl]{\(3\)}}

\put(09,22){\makebox(0,0)[bl]{\(2\)}}
\put(14,22){\makebox(0,0)[bl]{\(3\)}}
\put(19,22){\makebox(0,0)[bl]{\(2\)}}
\put(24,22){\makebox(0,0)[bl]{\(2\)}}
\put(29,22){\makebox(0,0)[bl]{\(2\)}}
\put(34,22){\makebox(0,0)[bl]{\(2\)}}
\put(39,22){\makebox(0,0)[bl]{\(3\)}}

\put(09,18){\makebox(0,0)[bl]{\(3\)}}
\put(14,18){\makebox(0,0)[bl]{\(3\)}}
\put(19,18){\makebox(0,0)[bl]{\(0\)}}
\put(24,18){\makebox(0,0)[bl]{\(0\)}}
\put(29,18){\makebox(0,0)[bl]{\(0\)}}
\put(34,18){\makebox(0,0)[bl]{\(1\)}}
\put(39,18){\makebox(0,0)[bl]{\(3\)}}

\put(29,14){\makebox(0,0)[bl]{\(3\)}}
\put(34,14){\makebox(0,0)[bl]{\(3\)}}
\put(39,14){\makebox(0,0)[bl]{\(3\)}}

\put(29,10){\makebox(0,0)[bl]{\(3\)}}
\put(34,10){\makebox(0,0)[bl]{\(3\)}}
\put(39,10){\makebox(0,0)[bl]{\(2\)}}

\put(29,06){\makebox(0,0)[bl]{\(1\)}}
\put(34,06){\makebox(0,0)[bl]{\(3\)}}
\put(39,06){\makebox(0,0)[bl]{\(3\)}}

\put(29,02){\makebox(0,0)[bl]{\(0\)}}
\put(34,02){\makebox(0,0)[bl]{\(3\)}}
\put(39,02){\makebox(0,0)[bl]{\(3\)}}

\end{picture}
\end{center}

   Fig. 4 depicts a personal computer
 (priorities of DAs are depicted in  parentheses,
 \(1\) corresponds to the best level;
 here the priorities are based on expert judgment):

 {\bf 0.} Notebook \(S\).

 {\bf 1.} Hardware \(H = B \star U \star V \star J\):


       {\it 1.1.} Mother board \(B\):~
      \(B_{1}\),
      \(B_{2}\);

      {\it 1.2.} CPU \(U\):~
      \(U_{1}\),
      \(U_{2}\),
      \(U_{3}\);

      {\it 1.3.} RAM \(E\):~
      \(E_{1}\),
      \(E_{2}\),
      \(E_{3}\),
      \(E_{4}\);

     {\it 1.4.} Hard drive \(V\):~
      \(V_{1}\),
      \(V_{2}\);

     {\it 1.5.} Video/graphic cards \(J\):~
      \(J_{1}\),
      \(J_{2}\).

 {\bf 2.} Software \(W = O \star D \star A \star G \):

   {\it 2.1.} Operation system OS \(O\):~
          \(O_{1}\),
          \(O_{2}\),
         \(O_{3}\);


%
%
%

  {\it 2.2.} Internet access (browser) \(A\):~
   \(A_{1}\),
   \(A_{2}\),
   \(A_{3}\),
   \(A_{4} =A_{2}\&A_{3} \);

%
  {\it 2.3.} Information processing (e.g., engineering
         software) \(G\):~
    \(G_{1}\),
   \(G_{2}\).
%


\begin{center}
\begin{picture}(78,76)
\put(19,0){\makebox(0,0)[bl] {Fig. 4. Personal computer}}


\put(00,72){\circle*{3}}

\put(03,70){\makebox(0,0)[bl]{Computer \(S = H \star W  \) }}

\put(00,64){\line(0,1){8}}


\put(30,65){\line(-1,0){30}}

\put(32,61){\makebox(0,0)[bl]{Software}}

\put(32,57){\makebox(0,0)[bl]{\(W = O  \star A \star G\)}}


\put(30,60){\circle*{2}} \put(30,57){\line(0,1){8}}

\put(30,53){\line(1,0){40}}


\put(30,48){\line(0,1){10}}


\put(47,48){\line(0,1){05}} \put(70,48){\line(0,1){05}}

\put(30,48){\circle*{1}}


\put(47,48){\circle*{1}} \put(70,48){\circle*{1}}


\put(32,49){\makebox(0,0)[bl]{\(O\)}}
\put(49,49){\makebox(0,0)[bl]{\(A\)}}
\put(72,49){\makebox(0,0)[bl]{\(G\)}}

\put(70,44){\makebox(0,0)[bl]{\(G_{1}(2)\)}}
\put(70,40){\makebox(0,0)[bl]{\(G_{2}(1)\)}}

\put(47,44){\makebox(0,0)[bl]{\(A_{1}(1)\)}}
\put(47,40){\makebox(0,0)[bl]{\(A_{2}(2)\)}}
\put(47,36){\makebox(0,0)[bl]{\(A_{3}(3)\)}}
\put(47,32){\makebox(0,0)[bl]{\(A_{4}=A_{2}\&A_{3}(3)\)}}


\put(30,44){\makebox(0,0)[bl]{\(O_{1}(1)\)}}
\put(30,40){\makebox(0,0)[bl]{\(O_{2}(2)\)}}
\put(30,36){\makebox(0,0)[bl]{\(O_{3}(3)\)}}


\put(02,34){\makebox(0,0)[bl]{Hardware}}

\put(02,30){\makebox(0,0)[bl]{\(H = B \star U \star E \star V
\star J\)}}

\put(00,42){\line(0,1){30}}


\put(00,37){\line(0,1){05}} \put(00,33){\circle*{2}}
\put(00,30){\line(0,1){11}}

\put(00,26){\line(1,0){40}}


\put(00,21){\line(0,1){10}} \put(10,21){\line(0,1){05}}
\put(20,21){\line(0,1){05}} \put(30,21){\line(0,1){05}}
\put(40,21){\line(0,1){05}}

\put(00,21){\circle*{1}} \put(10,21){\circle*{1}}
\put(20,21){\circle*{1}} \put(30,21){\circle*{1}}
\put(40,21){\circle*{1}}


\put(02,22){\makebox(0,0)[bl]{\(B\)}}
\put(12,22){\makebox(0,0)[bl]{\(U\)}}
\put(22,22){\makebox(0,0)[bl]{\(E\)}}
\put(32,22){\makebox(0,0)[bl]{\(V\)}}
\put(42,22){\makebox(0,0)[bl]{\(J\)}}

\put(40,17){\makebox(0,0)[bl]{\(J_{1}(1)\)}}
\put(40,13){\makebox(0,0)[bl]{\(J_{2}(2)\)}}

\put(30,17){\makebox(0,0)[bl]{\(V_{1}(1)\)}}
\put(30,13){\makebox(0,0)[bl]{\(V_{2}(2)\)}}

\put(20,17){\makebox(0,0)[bl]{\(E_{1}(1)\)}}
\put(20,13){\makebox(0,0)[bl]{\(E_{2}(1)\)}}
\put(20,09){\makebox(0,0)[bl]{\(E_{3}(2)\)}}
\put(20,05){\makebox(0,0)[bl]{\(E_{4}(3)\)}}

\put(10,17){\makebox(0,0)[bl]{\(U_{1}(1)\)}}
\put(10,13){\makebox(0,0)[bl]{\(U_{2}(2)\)}}
\put(10,09){\makebox(0,0)[bl]{\(U_{3}(3)\)}}

\put(00,17){\makebox(0,0)[bl]{\(B_{1}(1)\)}}
\put(00,13){\makebox(0,0)[bl]{\(B_{2}(2)\)}}

\end{picture}
\end{center}

 Table 3 and Table 4
 contain ordinal estimates of compatibility between DAs
 for different product components
 (\(3\) corresponds to the best level of compatibility,
 \(0\) corresponds to incompatibility).

\begin{center}
\begin{picture}(62,58)

\put(15,54){\makebox(0,0)[bl]{Table 3. Compatibility}}

\put(00,0){\line(1,0){62}} \put(00,46){\line(1,0){62}}
\put(00,52){\line(1,0){62}}

\put(00,0){\line(0,1){52}} \put(07,0){\line(0,1){52}}
\put(62,0){\line(0,1){52}}

\put(01,42){\makebox(0,0)[bl]{\(B_{1}\)}}
\put(01,38){\makebox(0,0)[bl]{\(B_{2}\)}}
\put(01,34){\makebox(0,0)[bl]{\(U_{1}\)}}
\put(01,30){\makebox(0,0)[bl]{\(U_{2}\)}}
\put(01,26){\makebox(0,0)[bl]{\(U_{3}\)}}
\put(01,22){\makebox(0,0)[bl]{\(E_{1}\)}}
\put(01,18){\makebox(0,0)[bl]{\(E_{2}\)}}
\put(01,14){\makebox(0,0)[bl]{\(E_{3}\)}}
\put(01,10){\makebox(0,0)[bl]{\(E_{4}\)}}
\put(01,06){\makebox(0,0)[bl]{\(V_{1}\)}}
\put(01,02){\makebox(0,0)[bl]{\(V_{2}\)}}

\put(12,46){\line(0,1){6}} \put(17,46){\line(0,1){6}}
\put(22,46){\line(0,1){6}} \put(27,46){\line(0,1){6}}
\put(32,46){\line(0,1){6}} \put(37,46){\line(0,1){6}}
\put(42,46){\line(0,1){6}} \put(47,46){\line(0,1){6}}
\put(52,46){\line(0,1){6}} \put(57,46){\line(0,1){6}}

\put(07.4,48){\makebox(0,0)[bl]{\(U_{1}\)}}
\put(12.4,48){\makebox(0,0)[bl]{\(U_{2}\)}}
\put(17.4,48){\makebox(0,0)[bl]{\(U_{3}\)}}
\put(22.4,48){\makebox(0,0)[bl]{\(E_{1}\)}}
\put(27.4,48){\makebox(0,0)[bl]{\(E_{2}\)}}
\put(32.4,48){\makebox(0,0)[bl]{\(E_{3}\)}}
\put(37.4,48){\makebox(0,0)[bl]{\(E_{4}\)}}
\put(42.4,48){\makebox(0,0)[bl]{\(V_{1}\)}}
\put(47.4,48){\makebox(0,0)[bl]{\(V_{2}\)}}
\put(52.4,48){\makebox(0,0)[bl]{\(J_{1}\)}}
\put(57.4,48){\makebox(0,0)[bl]{\(J_{2}\)}}


\put(09,42){\makebox(0,0)[bl]{\(3\)}}
\put(14,42){\makebox(0,0)[bl]{\(2\)}}
\put(19,42){\makebox(0,0)[bl]{\(2\)}}
\put(24,42){\makebox(0,0)[bl]{\(3\)}}
\put(29,42){\makebox(0,0)[bl]{\(3\)}}
\put(34,42){\makebox(0,0)[bl]{\(3\)}}
\put(39,42){\makebox(0,0)[bl]{\(3\)}}
\put(44,42){\makebox(0,0)[bl]{\(3\)}}
\put(49,42){\makebox(0,0)[bl]{\(2\)}}
\put(54,42){\makebox(0,0)[bl]{\(3\)}}
\put(59,42){\makebox(0,0)[bl]{\(2\)}}

\put(09,38){\makebox(0,0)[bl]{\(2\)}}
\put(14,38){\makebox(0,0)[bl]{\(3\)}}
\put(19,38){\makebox(0,0)[bl]{\(3\)}}
\put(24,38){\makebox(0,0)[bl]{\(2\)}}
\put(29,38){\makebox(0,0)[bl]{\(3\)}}
\put(34,38){\makebox(0,0)[bl]{\(3\)}}
\put(39,38){\makebox(0,0)[bl]{\(3\)}}
\put(44,38){\makebox(0,0)[bl]{\(2\)}}
\put(49,38){\makebox(0,0)[bl]{\(3\)}}
\put(54,38){\makebox(0,0)[bl]{\(2\)}}
\put(59,38){\makebox(0,0)[bl]{\(3\)}}

\put(24,34){\makebox(0,0)[bl]{\(3\)}}
\put(29,34){\makebox(0,0)[bl]{\(3\)}}
\put(34,34){\makebox(0,0)[bl]{\(3\)}}
\put(39,34){\makebox(0,0)[bl]{\(3\)}}
\put(44,34){\makebox(0,0)[bl]{\(3\)}}
\put(49,34){\makebox(0,0)[bl]{\(3\)}}
\put(54,34){\makebox(0,0)[bl]{\(3\)}}
\put(59,34){\makebox(0,0)[bl]{\(3\)}}

\put(24,30){\makebox(0,0)[bl]{\(2\)}}
\put(29,30){\makebox(0,0)[bl]{\(3\)}}
\put(34,30){\makebox(0,0)[bl]{\(3\)}}
\put(39,30){\makebox(0,0)[bl]{\(3\)}}
\put(44,30){\makebox(0,0)[bl]{\(3\)}}
\put(49,30){\makebox(0,0)[bl]{\(3\)}}
\put(54,30){\makebox(0,0)[bl]{\(3\)}}
\put(59,30){\makebox(0,0)[bl]{\(3\)}}

\put(24,26){\makebox(0,0)[bl]{\(2\)}}
\put(29,26){\makebox(0,0)[bl]{\(3\)}}
\put(34,26){\makebox(0,0)[bl]{\(3\)}}
\put(39,26){\makebox(0,0)[bl]{\(3\)}}
\put(44,26){\makebox(0,0)[bl]{\(2\)}}
\put(49,26){\makebox(0,0)[bl]{\(3\)}}
\put(54,26){\makebox(0,0)[bl]{\(2\)}}
\put(59,26){\makebox(0,0)[bl]{\(3\)}}

\put(44,22){\makebox(0,0)[bl]{\(3\)}}
\put(49,22){\makebox(0,0)[bl]{\(2\)}}
\put(54,22){\makebox(0,0)[bl]{\(3\)}}
\put(59,22){\makebox(0,0)[bl]{\(2\)}}

\put(44,18){\makebox(0,0)[bl]{\(2\)}}
\put(49,18){\makebox(0,0)[bl]{\(3\)}}
\put(54,18){\makebox(0,0)[bl]{\(2\)}}
\put(59,18){\makebox(0,0)[bl]{\(3\)}}

\put(44,14){\makebox(0,0)[bl]{\(2\)}}
\put(49,14){\makebox(0,0)[bl]{\(3\)}}
\put(54,14){\makebox(0,0)[bl]{\(2\)}}
\put(59,14){\makebox(0,0)[bl]{\(3\)}}

\put(44,10){\makebox(0,0)[bl]{\(2\)}}
\put(49,10){\makebox(0,0)[bl]{\(3\)}}
\put(54,10){\makebox(0,0)[bl]{\(2\)}}
\put(59,10){\makebox(0,0)[bl]{\(3\)}}

\put(54,06){\makebox(0,0)[bl]{\(3\)}}
\put(59,06){\makebox(0,0)[bl]{\(2\)}}

\put(54,02){\makebox(0,0)[bl]{\(2\)}}
\put(59,02){\makebox(0,0)[bl]{\(3\)}}

\end{picture}
\end{center}

\begin{center}
\begin{picture}(37,42)
\put(2.5,38){\makebox(0,0)[bl]{Table 4. Compatibility}}

\put(00,0){\line(1,0){37}} \put(00,30){\line(1,0){37}}
\put(00,36){\line(1,0){37}}

\put(00,0){\line(0,1){36}} \put(07,0){\line(0,1){36}}
\put(37,0){\line(0,1){36}}

\put(01,26){\makebox(0,0)[bl]{\(O_{1}\)}}
\put(01,22){\makebox(0,0)[bl]{\(O_{2}\)}}
\put(01,18){\makebox(0,0)[bl]{\(O_{3}\)}}
\put(01,14){\makebox(0,0)[bl]{\(A_{1}\)}}
\put(01,10){\makebox(0,0)[bl]{\(A_{2}\)}}
\put(01,06){\makebox(0,0)[bl]{\(A_{3}\)}}
\put(01,02){\makebox(0,0)[bl]{\(A_{4}\)}}

\put(12,30){\line(0,1){6}} \put(17,30){\line(0,1){6}}
\put(22,30){\line(0,1){6}} \put(27,30){\line(0,1){6}}
\put(32,30){\line(0,1){6}}

\put(07.4,32){\makebox(0,0)[bl]{\(A_{1}\)}}
\put(12.4,32){\makebox(0,0)[bl]{\(A_{2}\)}}
\put(17.4,32){\makebox(0,0)[bl]{\(A_{3}\)}}
\put(22.4,32){\makebox(0,0)[bl]{\(A_{4}\)}}
\put(27.4,32){\makebox(0,0)[bl]{\(G_{1}\)}}
\put(32.4,32){\makebox(0,0)[bl]{\(G_{2}\)}}


\put(09,26){\makebox(0,0)[bl]{\(3\)}}
\put(14,26){\makebox(0,0)[bl]{\(3\)}}
\put(19,26){\makebox(0,0)[bl]{\(3\)}}
\put(24,26){\makebox(0,0)[bl]{\(3\)}}
\put(29,26){\makebox(0,0)[bl]{\(2\)}}
\put(34,26){\makebox(0,0)[bl]{\(2\)}}

\put(09,22){\makebox(0,0)[bl]{\(3\)}}
\put(14,22){\makebox(0,0)[bl]{\(3\)}}
\put(19,22){\makebox(0,0)[bl]{\(3\)}}
\put(24,22){\makebox(0,0)[bl]{\(3\)}}
\put(29,22){\makebox(0,0)[bl]{\(3\)}}
\put(34,22){\makebox(0,0)[bl]{\(3\)}}

\put(09,18){\makebox(0,0)[bl]{\(3\)}}
\put(14,18){\makebox(0,0)[bl]{\(3\)}}
\put(19,18){\makebox(0,0)[bl]{\(3\)}}
\put(24,18){\makebox(0,0)[bl]{\(3\)}}
\put(29,18){\makebox(0,0)[bl]{\(3\)}}
\put(34,18){\makebox(0,0)[bl]{\(3\)}}

\put(29,14){\makebox(0,0)[bl]{\(3\)}}
\put(34,14){\makebox(0,0)[bl]{\(3\)}}

\put(29,10){\makebox(0,0)[bl]{\(3\)}}
\put(34,10){\makebox(0,0)[bl]{\(3\)}}

\put(29,06){\makebox(0,0)[bl]{\(3\)}}
\put(34,06){\makebox(0,0)[bl]{\(3\)}}

\put(29,02){\makebox(0,0)[bl]{\(3\)}}
\put(34,02){\makebox(0,0)[bl]{\(3\)}}

\end{picture}
\end{center}

\section{Basic Frameworks
 }

 A simplified scheme for selection
 (e.g., search  and multicriteria selection)
 of a required product is depicted in Fig. 5.
 Here the product is considered
 as a whole system.

 Recently, many products have a complex configuration and buyer
 can often generate a product configuration that
 is more useful for him/her.
 In Fig. 6,
 a multi-selection scheme  with composition
 of the resultant composite product
  from its components
  is presented.

 Further, a multi-selection  scheme
 for selection of structured products  and
 an aggregation of the resultant aggregated
 product(s) is presented in Fig. 7.

\begin{center}
\begin{picture}(40,37)

\put(02,00){\makebox(0,0)[bl]{Fig. 5. Selection
 scheme}}


\put(0,29){\line(1,0){40}} \put(0,36){\line(1,0){40}}
\put(0,29){\line(0,1){07}} \put(40,29){\line(0,1){07}}

\put(1,31){\makebox(0,0)[bl]{Internet product catalogue }}

\put(20,29){\vector(0,-1){4}}


\put(0,18){\line(1,0){40}} \put(0,25){\line(1,0){40}}
\put(0,18){\line(0,1){07}} \put(40,18){\line(0,1){07}}

\put(0.5,18){\line(0,1){07}} \put(39.5,18){\line(0,1){07}}

\put(7,20){\makebox(0,0)[bl]{Selection process}}

\put(20,18){\vector(0,-1){04}}


\put(20,09.5){\oval(40,9)} \put(20,9.5){\oval(39,8)}

\put(4,9.5){\makebox(0,0)[bl]{Selected product(s) \(S\)}}
\put(3.5,6){\makebox(0,0)[bl]{(as a recommendation)}}

\end{picture}
\end{center}

\begin{center}
\begin{picture}(80,64)

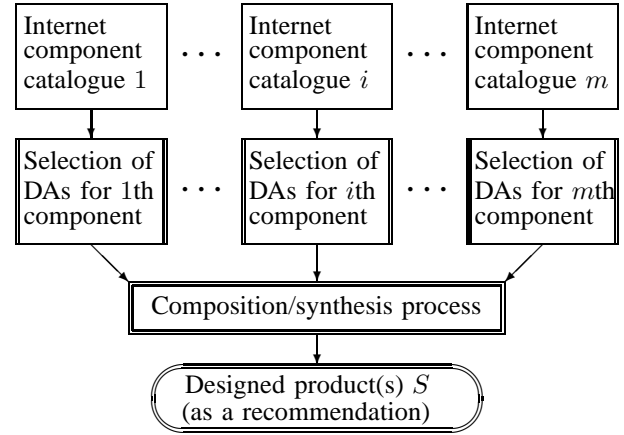
\put(06,00){\makebox(0,0)[bl]{Fig. 6. Multi-selection
 scheme (composition)
}}


\put(00,48){\line(1,0){20}} \put(00,62){\line(1,0){20}}
\put(00,48){\line(0,1){14}} \put(20,48){\line(0,1){14}}

\put(01,58){\makebox(0,0)[bl]{Internet}}
\put(01,54){\makebox(0,0)[bl]{component}}
\put(01,50){\makebox(0,0)[bl]{catalogue \(1\)}}

\put(10,48){\vector(0,-1){4}}


\put(22,55){\makebox(0,0)[bl]{{\bf . . .}}}
\put(52,55){\makebox(0,0)[bl]{{\bf . . .}}}


\put(30,48){\line(1,0){20}} \put(30,62){\line(1,0){20}}
\put(30,48){\line(0,1){14}} \put(50,48){\line(0,1){14}}

\put(31,58){\makebox(0,0)[bl]{Internet}}
\put(31,54){\makebox(0,0)[bl]{component }}
\put(31,50){\makebox(0,0)[bl]{catalogue \(i\)}}

\put(40,48){\vector(0,-1){4}}


\put(60,48){\line(1,0){20}} \put(60,62){\line(1,0){20}}
\put(60,48){\line(0,1){14}} \put(80,48){\line(0,1){14}}

\put(61,58){\makebox(0,0)[bl]{Internet}}
\put(61,54){\makebox(0,0)[bl]{component }}
\put(61,50){\makebox(0,0)[bl]{catalogue \(m\)}}

\put(70,48){\vector(0,-1){4}}



\put(00,30){\line(1,0){20}} \put(00,44){\line(1,0){20}}
\put(00,30){\line(0,1){14}} \put(20,30){\line(0,1){14}}

\put(0.4,30){\line(0,1){14}} \put(19.6,30){\line(0,1){14}}

\put(01,40){\makebox(0,0)[bl]{Selection of}}
\put(01,36){\makebox(0,0)[bl]{DAs for \(1\)th}}
\put(01,32){\makebox(0,0)[bl]{component }}

\put(10,30){\vector(1,-1){5}}


\put(22,37){\makebox(0,0)[bl]{{\bf . . .}}}
\put(52,37){\makebox(0,0)[bl]{{\bf . . .}}}


\put(30,30){\line(1,0){20}} \put(30,44){\line(1,0){20}}
\put(30,30){\line(0,1){14}} \put(50,30){\line(0,1){14}}

\put(30.4,30){\line(0,1){14}} \put(49.6,30){\line(0,1){14}}

\put(31,40){\makebox(0,0)[bl]{Selection of}}
\put(31,36){\makebox(0,0)[bl]{DAs for \(i\)th}}
\put(31,32){\makebox(0,0)[bl]{component}}

\put(40,30){\vector(0,-1){5}}


\put(60,30){\line(1,0){20}} \put(60,44){\line(1,0){20}}
\put(60,30){\line(0,1){14}} \put(80,30){\line(0,1){14}}

\put(60.4,30){\line(0,1){14}} \put(79.6,30){\line(0,1){14}}

\put(61,40){\makebox(0,0)[bl]{Selection of}}
\put(61,36){\makebox(0,0)[bl]{DAs for \(m\)th}}
\put(61,32){\makebox(0,0)[bl]{component}}

\put(70,30){\vector(-1,-1){5}}


\put(15,18){\line(1,0){50}} \put(15,25){\line(1,0){50}}
\put(15,18){\line(0,1){07}} \put(65,18){\line(0,1){07}}

\put(15.5,18.5){\line(1,0){49}} \put(15.5,24.5){\line(1,0){49}}
\put(15.5,18.5){\line(0,1){06}} \put(64.5,18.5){\line(0,1){06}}

\put(18,20){\makebox(0,0)[bl]{Composition/synthesis process}}

\put(40,18){\vector(0,-1){04}}


\put(40,09.5){\oval(44,9)} \put(40,9.5){\oval(43,8)}

\put(22.5,9.5){\makebox(0,0)[bl]{Designed product(s) \(S\)}}
\put(22,6){\makebox(0,0)[bl]{(as a recommendation)}}

\end{picture}
\end{center}

\begin{center}
\begin{picture}(80,84)

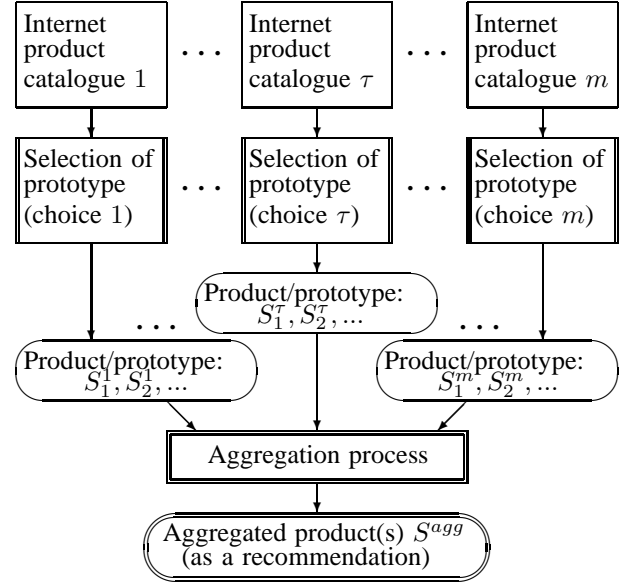
\put(08,00){\makebox(0,0)[bl]{Fig. 7. Multi-selection
 scheme
 (aggregation)
}}


\put(00,68){\line(1,0){20}} \put(00,82){\line(1,0){20}}
\put(00,68){\line(0,1){14}} \put(20,68){\line(0,1){14}}

\put(01,78){\makebox(0,0)[bl]{Internet}}
\put(01,74){\makebox(0,0)[bl]{product }}
\put(01,70){\makebox(0,0)[bl]{catalogue \(1\)}}

\put(10,68){\vector(0,-1){4}}


\put(22,75){\makebox(0,0)[bl]{{\bf . . .}}}
\put(52,75){\makebox(0,0)[bl]{{\bf . . .}}}


\put(30,68){\line(1,0){20}} \put(30,82){\line(1,0){20}}
\put(30,68){\line(0,1){14}} \put(50,68){\line(0,1){14}}

\put(31,78){\makebox(0,0)[bl]{Internet}}
\put(31,74){\makebox(0,0)[bl]{product }}
\put(31,70){\makebox(0,0)[bl]{catalogue \(\tau\)}}

\put(40,68){\vector(0,-1){4}}


\put(60,68){\line(1,0){20}} \put(60,82){\line(1,0){20}}
\put(60,68){\line(0,1){14}} \put(80,68){\line(0,1){14}}

\put(61,78){\makebox(0,0)[bl]{Internet}}
\put(61,74){\makebox(0,0)[bl]{product }}
\put(61,70){\makebox(0,0)[bl]{catalogue \(m\)}}

\put(70,68){\vector(0,-1){4}}



\put(00,50){\line(1,0){20}} \put(00,64){\line(1,0){20}}
\put(00,50){\line(0,1){14}} \put(20,50){\line(0,1){14}}

\put(0.4,50){\line(0,1){14}} \put(19.6,50){\line(0,1){14}}

\put(01,60){\makebox(0,0)[bl]{Selection of}}
\put(01,56){\makebox(0,0)[bl]{prototype}}
\put(01,52){\makebox(0,0)[bl]{(choice \(1\))}}

\put(10,50){\vector(0,-1){13}}


\put(22,57){\makebox(0,0)[bl]{{\bf . . .}}}
\put(52,57){\makebox(0,0)[bl]{{\bf . . .}}}


\put(30,50){\line(1,0){20}} \put(30,64){\line(1,0){20}}
\put(30,50){\line(0,1){14}} \put(50,50){\line(0,1){14}}

\put(30.4,50){\line(0,1){14}} \put(49.6,50){\line(0,1){14}}

\put(31,60){\makebox(0,0)[bl]{Selection of}}
\put(31,56){\makebox(0,0)[bl]{prototype}}
\put(31,52){\makebox(0,0)[bl]{(choice \(\tau\))}}

\put(40,50){\vector(0,-1){4}}


\put(60,50){\line(1,0){20}} \put(60,64){\line(1,0){20}}
\put(60,50){\line(0,1){14}} \put(80,50){\line(0,1){14}}

\put(60.4,50){\line(0,1){14}} \put(79.6,50){\line(0,1){14}}

\put(61,60){\makebox(0,0)[bl]{Selection of}}
\put(61,56){\makebox(0,0)[bl]{prototype}}
\put(61,52){\makebox(0,0)[bl]{(choice \(m\))}}

\put(70,50){\vector(0,-1){13}}


\put(16,33){\oval(32,08)}

\put(01,32.5){\makebox(0,0)[bl]{Product/prototype:}}
\put(09,29.5){\makebox(0,0)[bl]{\(S^{1}_{1},S^{1}_{2},...\)}}


\put(64,33){\oval(32,08)}

\put(49,32.5){\makebox(0,0)[bl]{Product/prototype:}}
\put(56,29.5){\makebox(0,0)[bl]{\(S^{m}_{1},S^{m}_{2},...\)}}


\put(40,42){\oval(32,8)}

\put(25,41.5){\makebox(0,0)[bl]{Product/prototype:}}
\put(32,38.5){\makebox(0,0)[bl]{\(S^{\tau}_{1},S^{\tau}_{2},...\)}}


\put(16,38.5){\makebox(0,0)[bl]{{\bf . . .}}}
\put(59,38.5){\makebox(0,0)[bl]{{\bf . . .}}}


\put(20,29){\vector(1,-1){4}} \put(40,38){\vector(0,-1){13}}
\put(60,29){\vector(-1,-1){4}}


\put(20,18){\line(1,0){40}} \put(20,25){\line(1,0){40}}
\put(20,18){\line(0,1){07}} \put(60,18){\line(0,1){07}}

\put(20.5,18.5){\line(1,0){39}} \put(20.5,24.5){\line(1,0){39}}
\put(20.5,18.5){\line(0,1){06}} \put(59.5,18.5){\line(0,1){06}}

\put(25.5,20){\makebox(0,0)[bl]{Aggregation process}}

\put(40,18){\vector(0,-1){04}}


\put(40,9.5){\oval(46,9)} \put(40,9.5){\oval(45,8)}

\put(19.5,9.5){\makebox(0,0)[bl]{Aggregated product(s)
\(S^{agg}\)}}

\put(22,6.5){\makebox(0,0)[bl]{(as a recommendation)}}

\end{picture}
\end{center}

 Clearly,
 multi-selection scheme with composition of product from its
 components
 and scheme of aggregation of selected modular products
 can be integrated into a resultant scheme:
 (i) selection of product components,
 (ii) synthesis of several modular products/prototypes, and
 (iii) aggregation of the obtained modular solutions into  the
 aggregated solution.


\section{Underlying Methods}



 The problem
 of multicriteria ranking (sorting problem)
 is the following
  (\cite{lev06}, \cite{zap02}).
 Let \(\Psi=\{1,...,i,...,p\}\) be a set of items which are evaluated upon
 criteria \( K = \{ 1,...,j,...,d \} \) and \(z_{i,j}\) is an estimate
 (quantitative, ordinal) of item \(i\) on criterion \(j\).
 The matrix \(\{z_{i,j}\}\)
 can be used as a basis to obtain a partial order on \(\Psi\)
 (i.e.,
  the following partition as linear ordered subsets of
 \(\Psi\)):
 \[\Psi= \cup_{k=1}^{m} \Psi(k),~ | \Psi(k_{1}) \cap \Psi(k_{2})| = 0~~
 if~~ k_{1}\neq k_{2},\]
 \[i_{2}\preceq i_{1}~~ \forall i_{1} \in \Psi(k_{1}),
 ~\forall i_{2} \in \Psi(k_{2}), ~ k_{1} \leq k_{2}.\]
 Set \(\Psi(k)\) is called layer \(k\), and each item \(i \in \Psi\)
 gets priority \(r_{i}\) that equals the number of the corresponding layer.
%
 In the paper, an outranking technique is used
 (\cite{levmih88}, \cite{roy96}).



The basic knapsack problem is
 (e.g., \cite{gar79}, \cite{kellerer04}, \cite{mar90}):
 \[\max\sum_{i=1}^{m} c_{i} x_{i}
 ~~ s.t.\sum_{i=1}^{m} a_{i} x_{i} \leq b,
 ~~ x_{i} \in\{0,1\}, ~ i=\overline{1,m}\]
 and additional resource constraints
 \[\sum_{i=1}^{m}a_{i,k} x_{i} \leq b_{k}; ~ k=\overline{1,l};\]
 where \(x_{i}=1\) if item \(i\) is selected,
 for \(i\)th item \(c_{i}\) is a value ('utility'), and
 \(a_{i}\) is a weight (i.e., resource requirement).
 Often nonnegative coefficients are assumed.
%
 In the case of  multiple choice problem,
 the items are divided into groups
 and
 it is necessary to select elements  (items)
 or the only one element
 from each group
 while taking into account a total resource constraint (or constraints):
 \[\max\sum_{i=1}^{m} \sum_{j=1}^{q_{i}} c_{ij} x_{ij}
 ~~ s.t.~\sum_{i=1}^{m} \sum_{j=1}^{q_{i}} a_{ij} x_{ij} \leq b,\]
 \[\sum_{j=1}^{q_{i}} x_{ij}=1,~ i=\overline{1,m};
 ~~ x_{ij} \in \{0, 1\}.\]
%
%
%
 The knapsack-like problems above
  are NP-hard
 and can be solved by the following approaches
 (\cite{gar79}, \cite{mar90}):
 (i) enumerative methods
 (e.g., Branch-and-Bound, dynamic programming),
 (ii) fully polynomial approximate schemes, and
 (iii) heuristics (e.g., greedy algorithms).
%
 In the paper, a greedy algorithm is used.


 Further,
  Hierarchical Morphological Multicriteria Design (HMMD)
 approach based on morphological clique problem is
 briefly described
  (\cite{lev98},  \cite{lev05}, \cite{lev06}).
%
 A examined composite
 (modular, decomposable) system consists
 of components and their interconnection or compatibility ({\bf I}).
 Basic assumptions of HMMD are the following:
 ~{\it (a)} a tree-like structure of the system;
 ~{\it (b)} a composite estimate for system quality
     that integrates components (subsystems, parts) qualities and
    qualities of IC (compatibility) across subsystems;
 ~{\it (c)} monotonic criteria for the system and its components;
 ~{\it (d)} quality of system components and {\bf I} are evaluated on the basis
    of coordinated ordinal scales.
 The designations are:
  ~(1) design alternatives (DAs) for leaf nodes of the model;
  ~(2) priorities of DAs (\(r=\overline{1,k}\);
      \(1\) corresponds to the best one);
  ~(3) ordinal compatibility ({\bf I}) for each pair of DAs
  (\(w=\overline{1,l}\); \(l\) corresponds to the best one).
 The basic phases of HMMD are:
  ~{\it 1.} design of the tree-like system model;
  ~{\it 2.} generation of DAs for leaf nodes of the model;
  ~{\it 3.} hierarchical selection and composing of DAs into composite
    DAs for the corresponding higher level of the system hierarchy;
  ~{\it 4.} analysis and improvement of composite DAs (decisions).

%

 Let \(S\) be a system consisting of \(m\) parts (components):
 \(P(1),...,P(i),...,P(m)\).
 A set of design alternatives
 is generated for each system part above.
 The problem is:


 {\it Find a composite design alternative}
 ~~ \(S=S(1)\star ...\star S(i)\star ...\star S(m)\)~~
 {\it of DAs (one representative design alternative}
 ~\(S(i)\)
 {\it for each system component/part}
  ~\(P(i)\), \(i=\overline{1,m}\)
  {\it )}
 {\it with non-zero}
 {\bf I}
 {\it between design alternatives.}


 A discrete space of the system excellence on the basis of the
 following vector is used:
 ~~\(N(S)=(w(S);n(S))\),
 ~where \(w(S)\) is the minimum of pairwise compatibility
 between DAs which correspond to different system components
 (i.e.,
 \(~\forall ~P_{j_{1}}\) and \( P_{j_{2}}\),
 \(1 \leq j_{1} \neq j_{2} \leq m\))
 in \(S\),
 ~\(n(S)=(n_{1},...,n_{r},...n_{k})\),
 ~where \(n_{r}\) is the number of DAs of the \(r\)th quality in \(S\).

 As a result,
 we search for composite decisions which are nondominated by \(N(S)\)
 (i.e., Pareto-efficient solutions).
%
%
 The considered combinatorial problem is NP-hard
 and an enumerative scheme is used.
 Fig. 8 and Fig. 9
 illustrate the composition problem by a numerical example
 (estimates of compatibility are pointed out in Fig. 9).
 In the example, composite DA is:
~\(S_{1}=X_{2}\star Y_{1}\star Z_{2}\), ~\(N(S_{1}) = (2;2,0,1)\).

\begin{center}
\begin{picture}(43,35)
\put(05,0){\makebox(0,0)[bl] {Fig. 8. Composition
 }}

\put(3,5){\makebox(0,8)[bl]{\(X_{3}(1)\)}}
\put(3,10){\makebox(0,8)[bl]{\(X_{2}(1)\)}}
\put(3,15){\makebox(0,8)[bl]{\(X_{1}(2)\)}}


\put(18,5){\makebox(0,8)[bl]{\(Y_{3}(2)\)}}
\put(18,10){\makebox(0,8)[bl]{\(Y_{2}(1)\)}}
\put(18,15){\makebox(0,8)[bl]{\(Y_{1}(3)\)}}

\put(33,5){\makebox(0,8)[bl]{\(Z_{3}(2)\)}}
\put(33,10){\makebox(0,8)[bl]{\(Z_{2}(1)\)}}
\put(33,15){\makebox(0,8)[bl]{\(Z_{1}(1)\)}}

\put(3,21){\circle*{2}} \put(18,21){\circle*{2}}
\put(33,21){\circle*{2}}

\put(0,21){\line(1,0){02}} \put(15,21){\line(1,0){02}}
\put(30,21){\line(1,0){02}}

\put(0,21){\line(0,-1){13}} \put(15,21){\line(0,-1){13}}
\put(30,21){\line(0,-1){13}}

\put(30,16){\line(1,0){01}} \put(30,12){\line(1,0){01}}
\put(30,8){\line(1,0){01}}

\put(32,16){\circle{1.7}} \put(32,16){\circle*{1}}
\put(32,12){\circle{1.7}} \put(32,12){\circle*{1}}
\put(32,8){\circle{1.7}} \put(32,8){\circle*{1}}

\put(15,8){\line(1,0){01}} \put(15,12){\line(1,0){01}}
\put(15,16){\line(1,0){01}}

\put(17,12){\circle{1.7}} \put(17,12){\circle*{1}}
\put(17,8){\circle{1.7}} \put(17,16){\circle{1.7}}
\put(17,8){\circle*{1}} \put(17,16){\circle*{1}}

\put(0,8){\line(1,0){01}} \put(0,12){\line(1,0){01}}
\put(0,16){\line(1,0){01}}

\put(2,12){\circle{1.7}} \put(2,16){\circle{1.7}}
\put(2,12){\circle*{1}} \put(2,16){\circle*{1}}
\put(2,8){\circle{1.7}} \put(2,8){\circle*{1}}

\put(3,26){\line(0,-1){04}} \put(18,26){\line(0,-1){04}}
\put(33,26){\line(0,-1){04}}


\put(3,26){\line(1,0){30}}


\put(07,26){\line(0,1){6}} \put(07,32){\circle*{3}}

\put(04,23){\makebox(0,8)[bl]{\(X\) }}
\put(14,23){\makebox(0,8)[bl]{\(Y\) }}
\put(29,23){\makebox(0,8)[bl]{\(Z\) }}

\put(11,32){\makebox(0,8)[bl]{\(S = X \star Y \star Z \) }}

\put(09,27){\makebox(0,8)[bl]{\(S_{1}=X_{2}\star Y_{1}\star
Z_{2}\)}}

\end{picture}
%
%
\begin{picture}(43,38)
\put(0,0){\makebox(0,0)[bl] {Fig. 9. Concentric presentation}}

\put(1,10){\line(0,1){6}} \put(2,10){\line(0,1){6}}

\put(01,10){\line(1,0){18}} \put(01,16){\line(1,0){18}}

\put(3,12){\makebox(0,0)[bl]{\(Z_{3}\)}} \put(8,10){\line(0,1){6}}
\put(9,12){\makebox(0,0)[bl]{\(Z_{2}\)}}
\put(14,12){\makebox(0,0)[bl]{\(Z_{1}\)}}
\put(19,10){\line(0,1){6}}

\put(25,10){\line(1,0){18}} \put(25,16){\line(1,0){18}}
\put(25,10){\line(0,1){6}}
\put(26,12){\makebox(0,0)[bl]{\(Y_{2}\)}}
\put(31,10){\line(0,1){6}}
\put(32,12){\makebox(0,0)[bl]{\(Y_{3}\)}}
\put(37,10){\line(0,1){6}}
\put(38,12){\makebox(0,0)[bl]{\(Y_{1}\)}}
\put(43,10){\line(0,1){6}}

\put(19,20){\line(0,1){18}} \put(25,20){\line(0,1){18}}
\put(19,20){\line(1,0){6}}
\put(20,21.5){\makebox(0,0)[bl]{\(X_{3}\)}}
\put(20,26){\makebox(0,0)[bl]{\(X_{2}\)}}
\put(19,30){\line(1,0){6}}
\put(20,32){\makebox(0,0)[bl]{\(X_{1}\)}}
\put(19,36){\line(1,0){6}} \put(19,38){\line(1,0){6}}

\put(10,10){\line(0,-1){3}} \put(10,7){\line(1,0){30}}
\put(40,7){\line(0,1){3}}

\put(07,6){\makebox(0,0)[bl]{\(2\)}}


\put(10,16){\line(0,1){11}} \put(10,27){\line(1,0){09}}
\put(11,22){\makebox(0,0)[bl]{\(2\)}}

\put(40,16){\line(0,1){11}} \put(40,27){\line(-1,0){15}}
\put(37,22){\makebox(0,0)[bl]{\(3\)}}


\end{picture}
\end{center}

 Aggregation of composite products (as modular solutions)
 can be considered as follows
 \cite{lev11agg}.
 Fig. 10 illustrates substructure, superstructure and
 ``kernel''
  (as a part of substructure)
 for three initial solutions
 \(S^{1}\), \(S^{2}\), and \(S^{3}\).

\begin{center}
\begin{picture}(58,28)
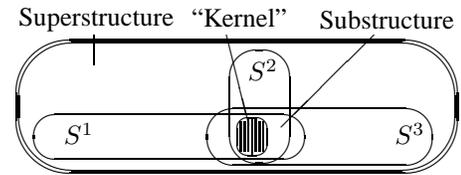
\put(00,0){\makebox(0,0)[bl]{Fig. 10. Substructure and
superstructure}}


\put(20,09){\oval(36,6)}

\put(06,8){\makebox(0,8)[bl]{\(S^{1}\)}}


\put(40,09){\oval(30,7.6)}

\put(50,8){\makebox(0,8)[bl]{\(S^{3}\)}}


\put(32,13){\oval(08,15)}

\put(30.5,16.5){\makebox(0,8)[bl]{\(S^{2}\)}}


\put(29,13){\oval(58,17)} \put(29,13){\oval(58.7,17.7)}

\put(00,23){\makebox(0,8)[bl]{Superstructure}}
 \put(10,22.5){\line(0,-1){4}}

 \put(23,23.6){\makebox(0,8)[bl]{``Kernel''}}
 \put(27.6,22.5){\line(1,-4){3}}

\put(40,23.4){\makebox(0,8)[bl]{Substructure}}
 \put(47.2,22.6){\line(-1,-1){12.3}}

\put(31,09){\oval(4,5)}

\put(29.5,7){\line(0,1){4}} \put(30,7){\line(0,1){4}}
\put(30.5,7){\line(0,1){4.5}} \put(31,6.5){\line(0,1){5}}
\put(31.5,7){\line(0,1){4.5}} \put(32,7){\line(0,1){4}}
\put(32.5,7){\line(0,1){4}}

\end{picture}
\end{center}

 In  \cite{lev11agg},
  basic aggregation strategies are described, for example:


 {\bf 1.} {\it Extension strategy}:
 ~{\it 1.1.} building a ``kernel'' for initial solutions
 (i.e., substrcuture/subsolution or an extended subsolution),
 ~{\it 1.2.} generation of a set of additional solution elements,
 ~{\it 1.3.} selection of additional elements from the generated
 set while taking into account their ``profit'' and resource requirements
 (i.e., a total ``profit'' and total resource constrain)
 (here knapsack-like problem is used).


  {\bf 2.} {\it Compression strategy}:
 ~{\it 2.1.} building a supersolution (as a superstructure),
 ~{\it 2.2.} generation of a set of solution elements from the built
 supersolution as candidates for deletion,
 ~{\it 2.3.} selection of the  elements-candidates for deletion
 while taking into account their ``profit'' and resource requirements
 (i.e., a total profit and total resource constrain)
 (here knapsack-like problem with minimization of objective function is used).


 Note, a general aggregation strategy has to be based on searching for a
 consensus/median solution
 \(S^{M}\) (``generalized'' median)
 for the initial solutions \( \overline{S} = \{S^{1},...,S^{n}\}\)
 (e.g.,  \cite{lev11agg}):
 \[ S^{M} = \arg ~ min_{X \in \overline{S}}  ~
 ( \sum_{i=1}^{n} ~ \rho (X, S^{i})  ),\]
  where
 \(\rho (X, Y)\) is a proximity (e.g., distance) between two solutions \(X\) and
 \(Y\).
  Mainly,
  searching for the median for many structures is usually NP complete
  problem.
 In our case, product structures correspond to a combination of
 tree, set of DAs, their estimates, matrices of compatibility
 estimates. As a result, the proximity between the structures
 are more complicated and
 the
 ``generalized'' median problem is very complex.
  Thus, simplified (approximate) solving strategies are often examined, for example
   \cite{lev11agg}:
  (a) searching for
  ``set median'' (i.e., one of the initial solutions is selected),
  (b) ``extension strategy'' above,
  (c) ``compression strategy'' above.

 {\bf 3.} {\it New design strategy}:
 {\it 3.1.} building a supersolution (as a superstructure)
 and design,
 {\it 3.2.} generation of  a ``design space''
 (as a product structure and design elements),
 {\it 3.3.} design of the composite solution over
  the obtained design element
  (here multiple choice problem or hierarchical morphological design approach can be used).

\section{Examples}

\subsection{Multicriteria Ranking/Selection}

 Multicriteria comparison/selection of product is
 the basic problem in multicriteria decision making.
%
 Table 5 contains an illustrative comparison example for five
 products,
  used criteria are (here
 ordinal scale is \([1,5]\),
  \(-\) corresponds to the case when minimum  estimate is the best,
 \(+\) corresponds to the case when maximum estimate is the best):
 cost \(K_{1}, -\),
 reliability \(K_{2}, +\),
 maintenance-ability \(K_{3} +\),
 upgrade-ability \(K_{3} +\).
 Evidently,
 two alternatives (products) \(A_{1}\) and \(A_{4}\)
 are Pareto-efficient solutions
 (corresponding priority equals \(1\)),
 alternative \(A_{2}\) is dominated by all others (priority equals
 \(3\)),
 and
  two alternatives \(A_{3}\) and \(A_{5}\)
  are intermediate by their quality
 (priority equals \(2\)).

\begin{center}
\begin{picture}(59,41)

\put(15,37){\makebox(0,0)[bl]{Table 5. Estimates}}

\put(00,00){\line(1,0){59}} \put(00,23){\line(1,0){59}}
\put(30,29){\line(1,0){16}} \put(00,35){\line(1,0){59}}

\put(00,00){\line(0,1){35}} \put(30,00){\line(0,1){35}}
\put(46,00){\line(0,1){35}} \put(59,00){\line(0,1){35}}

\put(34,23){\line(0,1){6}} \put(38,23){\line(0,1){6}}
\put(42,23){\line(0,1){6}}

\put(01,31){\makebox(0,0)[bl]{DAs}}

\put(32,31){\makebox(0,0)[bl]{Criteria}}

\put(47,31){\makebox(0,0)[bl]{Priority}}
\put(51.5,27){\makebox(0,0)[bl]{\(r_{i}\)}}

\put(31,25){\makebox(0,0)[bl]{\(1\)}}
\put(35,25){\makebox(0,0)[bl]{\(2\)}}
\put(39,25){\makebox(0,0)[bl]{\(3\)}}
\put(43,25){\makebox(0,0)[bl]{\(4\)}}

\put(01,18){\makebox(0,0)[bl]{\(A_{1}\) (computer \(1\))}}
\put(01,14){\makebox(0,0)[bl]{\(A_{2}\) (computer \(2\))}}
\put(01,10){\makebox(0,0)[bl]{\(A_{3}\) (computer \(3\))}}
\put(01,06){\makebox(0,0)[bl]{\(A_{4}\) (computer \(4\))}}
\put(01,02){\makebox(0,0)[bl]{\(A_{5}\) (computer \(5\))}}

\put(31,18){\makebox(0,0)[bl]{\(2\)}}
\put(31,14){\makebox(0,0)[bl]{\(3\)}}
\put(31,10){\makebox(0,0)[bl]{\(2\)}}
\put(31,06){\makebox(0,0)[bl]{\(1\)}}
\put(31,02){\makebox(0,0)[bl]{\(3\)}}

\put(35,18){\makebox(0,0)[bl]{\(4\)}}
\put(35,14){\makebox(0,0)[bl]{\(2\)}}
\put(35,10){\makebox(0,0)[bl]{\(4\)}}
\put(35,06){\makebox(0,0)[bl]{\(5\)}}
\put(35,02){\makebox(0,0)[bl]{\(3\)}}

\put(39,18){\makebox(0,0)[bl]{\(5\)}}
\put(39,14){\makebox(0,0)[bl]{\(1\)}}
\put(39,10){\makebox(0,0)[bl]{\(3\)}}
\put(39,06){\makebox(0,0)[bl]{\(4\)}}
\put(39,02){\makebox(0,0)[bl]{\(4\)}}

\put(43,18){\makebox(0,0)[bl]{\(4\)}}
\put(43,14){\makebox(0,0)[bl]{\(2\)}}
\put(43,10){\makebox(0,0)[bl]{\(3\)}}
\put(43,06){\makebox(0,0)[bl]{\(5\)}}
\put(43,02){\makebox(0,0)[bl]{\(4\)}}


\put(51,18){\makebox(0,0)[bl]{\(1\)}}
\put(51,14){\makebox(0,0)[bl]{\(3\)}}
\put(51,10){\makebox(0,0)[bl]{\(2\)}}
\put(51,06){\makebox(0,0)[bl]{\(1\)}}
\put(51,02){\makebox(0,0)[bl]{\(2\)}}

\end{picture}
\end{center}

\subsection{Synthesis of Composite Product}

 Here a numerical example of combinatorial synthesis
 (morphological design)
 of a composite product
 for the simplified example of three part motor vehicle
 is considered
 (Fig. 3, Table 2).
 This example corresponds to implementation of
 multi-selection scheme for composition of product components
 (Fig. 6).
 The obtained Pareto-efficient
 solutions are the following (Fig. 11):

 \(S_{1} = A_{1} \star B_{1} \star C_{2} \),
 \(N(S_{1}) = (3;2,1,0)\);

 \(S_{2} = A_{1} \star B_{1} \star C_{1} \),
 \(N(S_{2}) = (2;3,0,0)\).

\begin{center}
\begin{picture}(54,64)
\put(03,01){\makebox(0,0)[bl]{Fig. 11. Space of system quality}}



\put(0,011){\line(0,1){40}} \put(0,011){\line(3,4){15}}
\put(0,051){\line(3,-4){15}}

\put(20,016){\line(0,1){40}} \put(20,016){\line(3,4){15}}
\put(20,056){\line(3,-4){15}}

\put(40,021){\line(0,1){40}} \put(40,021){\line(3,4){15}}
\put(40,061){\line(3,-4){15}}

\put(20,55){\circle*{2}}
\put(08,53){\makebox(0,0)[bl]{\(N(S_{2})\)}}

\put(42.5,53){\circle*{2}}
\put(40.5,47){\makebox(0,0)[bl]{\(N(S_{1})\)}}

\put(40,61){\circle*{1}} \put(40,61){\circle{3}}

\put(24,60){\makebox(0,0)[bl]{The ideal}}
\put(27.5,57){\makebox(0,0)[bl]{point}}


\put(00,8){\makebox(0,0)[bl]{\(w=1\)}}
\put(20,13){\makebox(0,0)[bl]{\(w=2\)}}
\put(40,18){\makebox(0,0)[bl]{\(w=3\)}}
\end{picture}
\end{center}

\subsection{Synthesis of Extended Composite Product \cite{lev08a}}

 Now an extended composite product in electronic shopping is
 examined including the composite product,
 way of payment, place of purchase, etc.
 The simplified structure of the extended composite product
 ({\it buying a motor vehicle}) is depicted in Fig. 12:
 {\it 1.} origin of a motor vehicle \(A\)
 (domestic \(A_{1}\) foreign \(A_{2}\));
 {\it 2.} configuration of a motor vehicle \(B\)
 (minimal \(B_{1}\) and maximal \(B_{2}\));
 {\it 3.} way of payment \(C\)
 (credit \(C_{1}\), cash \(C_{2}\), and hire-purchase
 \(C_{3}\));
 {\it 4.} place of purchase \(D\)
 (motor vehicle store \(D_{1}\), motor vehicles dealer \(D_{2}\), and
 directly from manufacturer \(D_{3}\)); and
 {\it 5.} level of amortization \(E\)
 (new \(E_{1}\), used \(E_{2}\)).

The following criteria are used
 ('+' corresponds to positive orientation of an ordinal scale as \([1,5]\) and
 '-' corresponds to the negative orientation of the scale):
 {\it (a)} cost \(K_{a1}\) (-),
  brand prestigiousness \(K_{a2}\) (+),
  useful life \(K_{a3}\) (+),
  need of maintenance \(K_{a4}\) (-),
  reliability \(K_{a5}\) (+);
 {\it (b)}  cost \(K_{b1}\) (-),
  brand prestigiousness \(K_{b2}\) (+),
  upgradeability \(K_{b3}\) (+);
  {\it (c)} credit risk \(K_{c1}\) (-),
  cost of usage \(K_{c2}\) (+),
  availability \(K_{c3}\) (+);
  {\it (d)} reliability \(K_{d1}\) (-),
  cost \(K_{d2}\) (+),
 service quality \(K_{d3}\) (+),
  warranty \(K_{d4}\) (-); and
  {\it (e)} cost \(K_{e1}\) (-),
  need of maintenance \(K_{e2}\) (+),
  warranty \(K_{e3}\) (+).

\begin{center}
\begin{picture}(72,40)
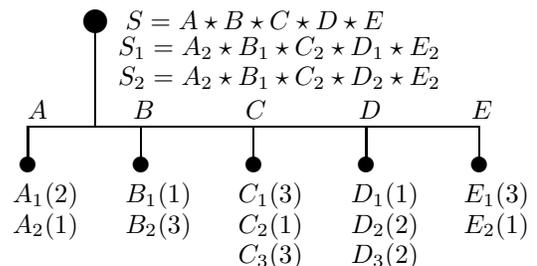
\put(06,00){\makebox(0,0)[bl] {Fig. 12. Structure of extended
product}}

\put(1,12){\makebox(0,8)[bl]{\(A_{1}(2)\)}}
\put(1,08){\makebox(0,8)[bl]{\(A_{2}(1)\)}}

\put(16,12){\makebox(0,8)[bl]{\(B_{1}(1)\)}}
\put(16,8){\makebox(0,8)[bl]{\(B_{2}(3)\)}}

\put(31,12){\makebox(0,8)[bl]{\(C_{1}(3)\)}}
\put(31,8){\makebox(0,8)[bl]{\(C_{2}(1)\)}}
\put(31,04){\makebox(0,8)[bl]{\(C_{3}(3)\)}}

\put(46,12){\makebox(0,8)[bl]{\(D_{1}(1)\)}}
\put(46,8){\makebox(0,8)[bl]{\(D_{2}(2)\)}}
\put(46,04){\makebox(0,8)[bl]{\(D_{3}(2)\)}}

\put(61,12){\makebox(0,8)[bl]{\(E_{1}(3)\)}}
\put(61,08){\makebox(0,8)[bl]{\(E_{2}(1)\)}}


\put(3,18){\circle*{2}} \put(18,18){\circle*{2}}
\put(33,18){\circle*{2}} \put(48,18){\circle*{2}}
\put(63,18){\circle*{2}}

\put(3,23){\line(0,-1){04}} \put(18,23){\line(0,-1){04}}
\put(33,23){\line(0,-1){04}} \put(48,23){\line(0,-1){04}}
\put(63,23){\line(0,-1){04}}


\put(3,23){\line(1,0){60}}

\put(03,24){\makebox(0,8)[bl]{\(A\) }}
\put(17,24){\makebox(0,8)[bl]{\(B\) }}
\put(32,24){\makebox(0,8)[bl]{\(C\) }}
\put(47,24){\makebox(0,8)[bl]{\(D\) }}
\put(62,24){\makebox(0,8)[bl]{\(E\) }}


\put(12,23){\line(0,1){14}} \put(12,37){\circle*{3}}

\put(16,36){\makebox(0,8)[bl] {\(S = A \star B \star C \star D
\star E \)}}

\put(15,32){\makebox(0,8)[bl] {\(S_{1} = A_{2} \star B_{1} \star
C_{2} \star D_{1} \star E_{2} \)}}

\put(15,28){\makebox(0,8)[bl] {\(S_{2} = A_{2} \star B_{1} \star
C_{2} \star D_{2} \star E_{2} \)}}

\end{picture}
\end{center}

 Tables 6, 7, 8, 9, and 10 contain ordinal estimates of DAs upon the
 above-mentioned criteria (expert judgment).
 Estimates of compatibility between DAs are contained
 in Table 11 (scale \([0,3]\), expert judgment).

\begin{center}
\begin{picture}(34,30)
\put(02,26){\makebox(0,0)[bl]{Table 6. Estimates}}

\put(00,03){\line(1,0){30}} \put(00,13){\line(1,0){30}}
\put(10,19){\line(1,0){20}} \put(00,25){\line(1,0){30}}

\put(00,03){\line(0,1){22}} \put(10,03){\line(0,1){22}}
\put(30,03){\line(0,1){22}}

\put(14,13){\line(0,1){6}} \put(18,13){\line(0,1){6}}
\put(22,13){\line(0,1){6}} \put(26,13){\line(0,1){6}}

\put(01,21){\makebox(0,0)[bl]{DAs}}
\put(14,21){\makebox(0,0)[bl]{Criteria}}

\put(11,15){\makebox(0,0)[bl]{\(1\)}}
\put(15,15){\makebox(0,0)[bl]{\(2\)}}
\put(19,15){\makebox(0,0)[bl]{\(3\)}}
\put(23,15){\makebox(0,0)[bl]{\(4\)}}
\put(27,15){\makebox(0,0)[bl]{\(5\)}}

\put(01,09){\makebox(0,0)[bl]{\(A_{1}\)}}
\put(01,05){\makebox(0,0)[bl]{\(A_{2}\)}}

\put(11,09){\makebox(0,0)[bl]{\(2\)}}
\put(11,05){\makebox(0,0)[bl]{\(4\)}}

\put(15,09){\makebox(0,0)[bl]{\(3\)}}
\put(15,05){\makebox(0,0)[bl]{\(5\)}}

\put(19,09){\makebox(0,0)[bl]{\(3\)}}
\put(19,05){\makebox(0,0)[bl]{\(5\)}}

\put(23,09){\makebox(0,0)[bl]{\(3\)}}
\put(23,05){\makebox(0,0)[bl]{\(5\)}}

\put(27,09){\makebox(0,0)[bl]{\(2\)}}
\put(27,05){\makebox(0,0)[bl]{\(4\)}}

\end{picture}
\begin{picture}(25,30)

\put(00,26){\makebox(0,0)[bl]{Table 7. Estimates}}

\put(00,03){\line(1,0){23}} \put(00,13){\line(1,0){23}}
\put(10,19){\line(1,0){13}} \put(00,25){\line(1,0){23}}

\put(00,03){\line(0,1){22}} \put(10,03){\line(0,1){22}}
\put(23,03){\line(0,1){22}}

\put(14,13){\line(0,1){6}} \put(18,13){\line(0,1){6}}

\put(01,21){\makebox(0,0)[bl]{DAs}}
\put(10.5,21){\makebox(0,0)[bl]{Criteria}}

\put(11,15){\makebox(0,0)[bl]{\(1\)}}
\put(15,15){\makebox(0,0)[bl]{\(2\)}}
\put(19,15){\makebox(0,0)[bl]{\(3\)}}

\put(01,09){\makebox(0,0)[bl]{\(B_{1}\)}}
\put(01,05){\makebox(0,0)[bl]{\(B_{2}\)}}

\put(11,09){\makebox(0,0)[bl]{\(2\)}}
\put(11,05){\makebox(0,0)[bl]{\(4\)}}

\put(15,09){\makebox(0,0)[bl]{\(3\)}}
\put(15,05){\makebox(0,0)[bl]{\(5\)}}

\put(19,09){\makebox(0,0)[bl]{\(5\)}}
\put(19,05){\makebox(0,0)[bl]{\(2\)}}

\end{picture}
\end{center}

\begin{center}
\begin{picture}(30,34)

\put(00,30){\makebox(0,0)[bl]{Table 8. Estimates}}

\put(00,03){\line(1,0){23}} \put(00,17){\line(1,0){23}}
\put(10,23){\line(1,0){13}} \put(00,29){\line(1,0){23}}

\put(00,03){\line(0,1){26}} \put(10,03){\line(0,1){26}}
\put(23,03){\line(0,1){26}}

\put(14,17){\line(0,1){6}} \put(18,17){\line(0,1){6}}

\put(01,25){\makebox(0,0)[bl]{DAs}}
\put(10.5,25){\makebox(0,0)[bl]{Criteria}}

\put(11,19){\makebox(0,0)[bl]{\(1\)}}
\put(15,19){\makebox(0,0)[bl]{\(2\)}}
\put(19,19){\makebox(0,0)[bl]{\(3\)}}

\put(01,13){\makebox(0,0)[bl]{\(C_{1}\)}}
\put(01,09){\makebox(0,0)[bl]{\(C_{2}\)}}
\put(01,05){\makebox(0,0)[bl]{\(C_{3}\)}}

\put(11,13){\makebox(0,0)[bl]{\(5\)}}
\put(11,09){\makebox(0,0)[bl]{\(1\)}}
\put(11,05){\makebox(0,0)[bl]{\(5\)}}

\put(15,13){\makebox(0,0)[bl]{\(5\)}}
\put(15,09){\makebox(0,0)[bl]{\(4\)}}
\put(15,05){\makebox(0,0)[bl]{\(3\)}}

\put(19,13){\makebox(0,0)[bl]{\(4\)}}
\put(19,09){\makebox(0,0)[bl]{\(3\)}}
\put(19,05){\makebox(0,0)[bl]{\(4\)}}

\end{picture}
\begin{picture}(26,34)

\put(00,30){\makebox(0,0)[bl]{Table 9. Estimates}}

\put(00,03){\line(1,0){26}} \put(00,17){\line(1,0){26}}
\put(10,23){\line(1,0){16}} \put(00,29){\line(1,0){26}}

\put(00,03){\line(0,1){26}} \put(10,03){\line(0,1){26}}
\put(26,03){\line(0,1){26}}

\put(14,17){\line(0,1){6}} \put(18,17){\line(0,1){6}}
\put(22,17){\line(0,1){6}}

\put(01,25){\makebox(0,0)[bl]{DAs}}
\put(12,25){\makebox(0,0)[bl]{Criteria}}

\put(11,19){\makebox(0,0)[bl]{\(1\)}}
\put(15,19){\makebox(0,0)[bl]{\(2\)}}
\put(19,19){\makebox(0,0)[bl]{\(3\)}}
\put(23,19){\makebox(0,0)[bl]{\(4\)}}

\put(01,13){\makebox(0,0)[bl]{\(D_{1}\)}}
\put(01,09){\makebox(0,0)[bl]{\(D_{2}\)}}
\put(01,05){\makebox(0,0)[bl]{\(D_{3}\)}}

\put(11,13){\makebox(0,0)[bl]{\(4\)}}
\put(11,09){\makebox(0,0)[bl]{\(2\)}}
\put(11,05){\makebox(0,0)[bl]{\(3\)}}

\put(15,13){\makebox(0,0)[bl]{\(4\)}}
\put(15,09){\makebox(0,0)[bl]{\(3\)}}
\put(15,05){\makebox(0,0)[bl]{\(3\)}}

\put(19,13){\makebox(0,0)[bl]{\(4\)}}
\put(19,09){\makebox(0,0)[bl]{\(2\)}}
\put(19,05){\makebox(0,0)[bl]{\(1\)}}

\put(23,13){\makebox(0,0)[bl]{\(5\)}}
\put(23,09){\makebox(0,0)[bl]{\(2\)}}
\put(23,05){\makebox(0,0)[bl]{\(2\)}}

\end{picture}
\end{center}

\begin{center}
\begin{picture}(23,30)

\put(00,26){\makebox(0,0)[bl]{Table 10. Estimates}}

\put(00,03){\line(1,0){23}} \put(00,13){\line(1,0){23}}
\put(10,19){\line(1,0){13}} \put(00,25){\line(1,0){23}}

\put(00,03){\line(0,1){22}} \put(10,03){\line(0,1){22}}
\put(23,03){\line(0,1){22}}

\put(14,13){\line(0,1){6}} \put(18,13){\line(0,1){6}}

\put(01,21){\makebox(0,0)[bl]{DAs}}
\put(10.5,21){\makebox(0,0)[bl]{Criteria}}

\put(11,15){\makebox(0,0)[bl]{\(1\)}}
\put(15,15){\makebox(0,0)[bl]{\(2\)}}
\put(19,15){\makebox(0,0)[bl]{\(3\)}}

\put(01,09){\makebox(0,0)[bl]{\(E_{2}\)}}
\put(01,05){\makebox(0,0)[bl]{\(E_{1}\)}}

\put(11,09){\makebox(0,0)[bl]{\(4\)}}
\put(11,05){\makebox(0,0)[bl]{\(2\)}}

\put(15,09){\makebox(0,0)[bl]{\(2\)}}
\put(15,05){\makebox(0,0)[bl]{\(4\)}}

\put(19,09){\makebox(0,0)[bl]{\(5\)}}
\put(19,05){\makebox(0,0)[bl]{\(1\)}}

\end{picture}
\end{center}

\begin{center}
\begin{picture}(57,53)

\put(12,49){\makebox(0,0)[bl]{Table 11. Compatibility}}

\put(00,0){\line(1,0){57}} \put(00,42){\line(1,0){57}}
\put(00,48){\line(1,0){57}}

\put(00,0){\line(0,1){48}} \put(07,0){\line(0,1){48}}
\put(57,0){\line(0,1){48}}

\put(01,38){\makebox(0,0)[bl]{\(A_{1}\)}}
\put(01,34){\makebox(0,0)[bl]{\(A_{2}\)}}
\put(01,30){\makebox(0,0)[bl]{\(B_{1}\)}}
\put(01,26){\makebox(0,0)[bl]{\(B_{2}\)}}

\put(01,22){\makebox(0,0)[bl]{\(C_{1}\)}}
\put(01,18){\makebox(0,0)[bl]{\(C_{2}\)}}
\put(01,14){\makebox(0,0)[bl]{\(C_{3}\)}}
\put(01,10){\makebox(0,0)[bl]{\(D_{1}\)}}
\put(01,06){\makebox(0,0)[bl]{\(D_{2}\)}}
\put(01,02){\makebox(0,0)[bl]{\(D_{3}\)}}

\put(12,42){\line(0,1){6}} \put(17,42){\line(0,1){6}}
\put(22,42){\line(0,1){6}} \put(27,42){\line(0,1){6}}
\put(32,42){\line(0,1){6}} \put(37,42){\line(0,1){6}}
\put(42,42){\line(0,1){6}} \put(47,42){\line(0,1){6}}
\put(52,42){\line(0,1){6}}

\put(07.4,44){\makebox(0,0)[bl]{\(B_{1}\)}}
\put(12.4,44){\makebox(0,0)[bl]{\(B_{2}\)}}
\put(17.4,44){\makebox(0,0)[bl]{\(C_{1}\)}}
\put(22.4,44){\makebox(0,0)[bl]{\(C_{2}\)}}
\put(27.4,44){\makebox(0,0)[bl]{\(C_{3}\)}}
\put(32.4,44){\makebox(0,0)[bl]{\(D_{1}\)}}
\put(37.4,44){\makebox(0,0)[bl]{\(D_{2}\)}}
\put(42.4,44){\makebox(0,0)[bl]{\(D_{3}\)}}
\put(47.4,44){\makebox(0,0)[bl]{\(E_{1}\)}}
\put(52.4,44){\makebox(0,0)[bl]{\(E_{2}\)}}

\put(09,38){\makebox(0,0)[bl]{\(3\)}}
\put(14,38){\makebox(0,0)[bl]{\(3\)}}
\put(19,38){\makebox(0,0)[bl]{\(2\)}}
\put(24,38){\makebox(0,0)[bl]{\(3\)}}
\put(29,38){\makebox(0,0)[bl]{\(2\)}}
\put(34,38){\makebox(0,0)[bl]{\(3\)}}
\put(39,38){\makebox(0,0)[bl]{\(3\)}}
\put(44,38){\makebox(0,0)[bl]{\(0\)}}
\put(49,38){\makebox(0,0)[bl]{\(3\)}}
\put(54,38){\makebox(0,0)[bl]{\(3\)}}

\put(09,34){\makebox(0,0)[bl]{\(3\)}}
\put(14,34){\makebox(0,0)[bl]{\(3\)}}
\put(19,34){\makebox(0,0)[bl]{\(3\)}}
\put(24,34){\makebox(0,0)[bl]{\(3\)}}
\put(29,34){\makebox(0,0)[bl]{\(3\)}}
\put(34,34){\makebox(0,0)[bl]{\(3\)}}
\put(39,34){\makebox(0,0)[bl]{\(3\)}}
\put(44,34){\makebox(0,0)[bl]{\(2\)}}
\put(49,34){\makebox(0,0)[bl]{\(3\)}}
\put(54,34){\makebox(0,0)[bl]{\(3\)}}

\put(19,30){\makebox(0,0)[bl]{\(3\)}}
\put(24,30){\makebox(0,0)[bl]{\(3\)}}
\put(29,30){\makebox(0,0)[bl]{\(3\)}}
\put(34,30){\makebox(0,0)[bl]{\(3\)}}
\put(39,30){\makebox(0,0)[bl]{\(3\)}}
\put(44,30){\makebox(0,0)[bl]{\(2\)}}
\put(49,30){\makebox(0,0)[bl]{\(3\)}}
\put(54,30){\makebox(0,0)[bl]{\(3\)}}

\put(19,26){\makebox(0,0)[bl]{\(3\)}}
\put(24,26){\makebox(0,0)[bl]{\(3\)}}
\put(29,26){\makebox(0,0)[bl]{\(3\)}}
\put(34,26){\makebox(0,0)[bl]{\(3\)}}
\put(39,26){\makebox(0,0)[bl]{\(2\)}}
\put(44,26){\makebox(0,0)[bl]{\(1\)}}
\put(49,26){\makebox(0,0)[bl]{\(3\)}}
\put(54,26){\makebox(0,0)[bl]{\(2\)}}

\put(34,22){\makebox(0,0)[bl]{\(3\)}}
\put(39,22){\makebox(0,0)[bl]{\(1\)}}
\put(44,22){\makebox(0,0)[bl]{\(0\)}}
\put(49,22){\makebox(0,0)[bl]{\(3\)}}
\put(54,22){\makebox(0,0)[bl]{\(1\)}}

\put(34,18){\makebox(0,0)[bl]{\(3\)}}
\put(39,18){\makebox(0,0)[bl]{\(3\)}}
\put(44,18){\makebox(0,0)[bl]{\(2\)}}
\put(49,18){\makebox(0,0)[bl]{\(3\)}}
\put(54,18){\makebox(0,0)[bl]{\(3\)}}

\put(34,14){\makebox(0,0)[bl]{\(2\)}}
\put(39,14){\makebox(0,0)[bl]{\(0\)}}
\put(44,14){\makebox(0,0)[bl]{\(0\)}}
\put(49,14){\makebox(0,0)[bl]{\(3\)}}
\put(54,14){\makebox(0,0)[bl]{\(0\)}}

\put(49,10){\makebox(0,0)[bl]{\(3\)}}
\put(54,10){\makebox(0,0)[bl]{\(1\)}}

\put(49,06){\makebox(0,0)[bl]{\(2\)}}
\put(54,06){\makebox(0,0)[bl]{\(3\)}}

\put(49,02){\makebox(0,0)[bl]{\(1\)}}
\put(54,02){\makebox(0,0)[bl]{\(3\)}}

\end{picture}
\end{center}

 The resultant priorities of DAs are obtained on the basis of
 multicriteria ranking for each system part (scale \([1,3]\)).
 The priorities are
 shown in Fig. 12 in parentheses.
%


 The resultant composite Pareto-efficient DAs are the following
 (Fig. 13):

 \(S_{1} = A_{2} \star B_{1} \star C_{2} \star D_{1} \star E_{2}\),
 \(N(S_{1})= (1;5,0,0)\) and

 \(S_{2} = A_{2} \star B_{1} \star C_{2} \star D_{2} \star E_{2}\),
 \(N(S_{2})= (3;4,1,0)\).


\begin{center}
\begin{picture}(56,64)
\put(03,01){\makebox(0,0)[bl]{Fig. 13. Space of system quality}}



\put(00,011){\line(0,1){40}} \put(00,011){\line(3,4){15}}
\put(00,051){\line(3,-4){15}}

\put(20,016){\line(0,1){40}} \put(20,016){\line(3,4){15}}
\put(20,056){\line(3,-4){15}}

\put(40,021){\line(0,1){40}} \put(40,021){\line(3,4){15}}
\put(40,061){\line(3,-4){15}}

\put(00,50){\circle*{2}}
\put(03,48){\makebox(0,0)[bl]{\(N(S_{1})\)}}

\put(42.5,53){\circle*{2}}
\put(40.5,47){\makebox(0,0)[bl]{\(N(S_{2})\)}}

\put(40,61){\circle*{1}} \put(40,61){\circle{3}}

\put(24,59){\makebox(0,0)[bl]{The ideal}}
\put(26.5,55){\makebox(0,0)[bl]{point}}


\put(00,8){\makebox(0,0)[bl]{\(w=1\)}}
\put(20,13){\makebox(0,0)[bl]{\(w=2\)}}
\put(40,18){\makebox(0,0)[bl]{\(w=3\)}}
\end{picture}
\end{center}

\subsection{Synthesis of Product Repair Plan \cite{lev08a}}

 For complex products it is often necessary to consider repair plans.
 The described example corresponds to a car.
%
 Generally, the car repair plan consists of the following parts:
 (1) payment,
 (2) body,
 (3) electric \& electronic subsystem, and
 (4) tuning, and
 (5) motor vehicle.
 Here a compressed plan is examined as follows (Fig. 14)
 (priorities of DAs are based on
 expert judgment and
 shown in parentheses):

\begin{center}
\begin{picture}(72,116)

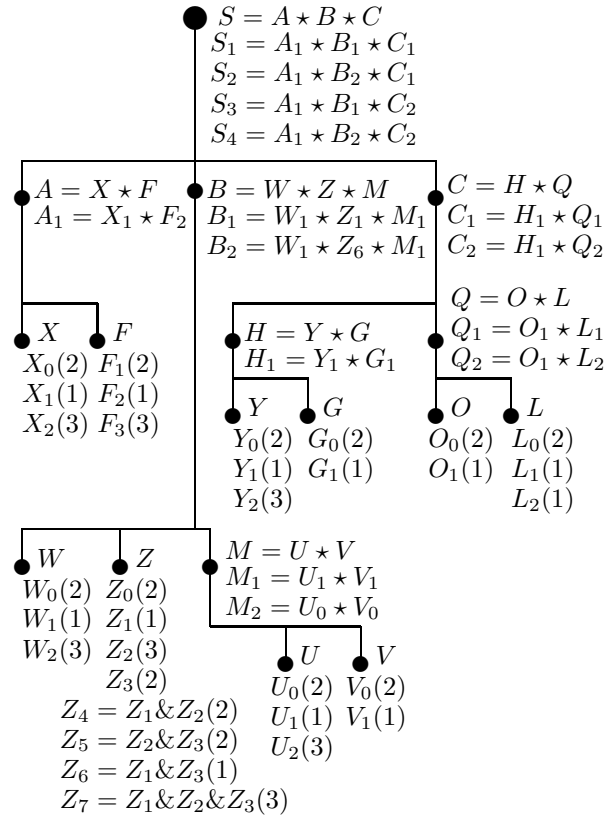
\put(12,0){\makebox(0,0)[bl] {Fig. 14. Structure of repair plan}}

\put(26,111){\makebox(0,8)[bl]{\(S=A\star B\star C\)}}

\put(25,107){\makebox(0,8)[bl]{\(S_{1}=A_{1}\star B_{1} \star
C_{1}\)}}

\put(25,103){\makebox(0,8)[bl]{\(S_{2}=A_{1}\star B_{2} \star
C_{1}\)}}

\put(25,99){\makebox(0,8)[bl]{\(S_{3}=A_{1}\star B_{1} \star
C_{2}\)}}

\put(25,95){\makebox(0,8)[bl]{\(S_{4}=A_{1}\star B_{2} \star
C_{2}\)}}

\put(23,112){\circle*{3.3}}

\put(23,93){\line(0,1){18}}

\put(0,93){\line(1,0){55}}


\put(23,90){\line(0,1){3}}

\put(24.5,88){\makebox(0,8)[bl]{\(B = W\star Z\star M\)}}

\put(24.5,84){\makebox(0,8)[bl]{\(B_{1} = W_{1}\star Z_{1} \star
M_{1}\)}}

\put(24.5,80){\makebox(0,8)[bl]{\(B_{2} = W_{1}\star Z_{6} \star
M_{1}\)}}


\put(23,89){\circle*{2}}

\put(23,44){\line(0,1){44}}


\put(0,89){\line(0,1){4}}

\put(1.5,88){\makebox(0,8)[bl]{\(A=X\star F\)}}

\put(1.5,84){\makebox(0,8)[bl]{\(A_{1}=X_{1}\star F_{2}\)}}

\put(0,88){\circle*{2}}

\put(0,74){\line(0,1){13}}


\put(0,74){\line(1,0){10}}

\put(0,70){\line(0,1){4}}

\put(0,69){\circle*{2}}

\put(2,69){\makebox(0,8)[bl]{\(X\)}}

\put(0,64){\makebox(0,8)[bl]{\(X_{0}(2)\)}}
\put(0,60){\makebox(0,8)[bl]{\(X_{1}(1)\)}}
\put(0,56){\makebox(0,8)[bl]{\(X_{2}(3)\)}}
%


\put(10,70){\line(0,1){4}}

\put(10,69){\circle*{2}}

\put(12,69){\makebox(0,8)[bl]{\(F\)}}

\put(10,64){\makebox(0,8)[bl]{\(F_{1}(2)\)}}
\put(10,60){\makebox(0,8)[bl]{\(F_{2}(1)\)}}
\put(10,56){\makebox(0,8)[bl]{\(F_{3}(3)\)}}


\put(0,44){\line(1,0){25}}

\put(0,40){\line(0,1){4}}

\put(0,39){\circle*{2}}

\put(2,39){\makebox(0,8)[bl]{\(W\)}}

\put(0,34){\makebox(0,8)[bl]{\(W_{0}(2)\)}}
\put(0,30){\makebox(0,8)[bl]{\(W_{1}(1)\)}}
\put(0,26){\makebox(0,8)[bl]{\(W_{2}(3)\)}}
%


\put(13,40){\line(0,1){4}}

\put(13,39){\circle*{2}}

\put(15,39){\makebox(0,8)[bl]{\(Z\)}}

\put(11,34){\makebox(0,8)[bl]{\(Z_{0}(2)\)}}
\put(11,30){\makebox(0,8)[bl]{\(Z_{1}(1)\)}}
\put(11,26){\makebox(0,8)[bl]{\(Z_{2}(3)\)}}
\put(11,22){\makebox(0,8)[bl]{\(Z_{3}(2)\)}}
\put(05,18){\makebox(0,8)[bl]{\(Z_{4}=Z_{1}\&Z_{2}(2)\)}}
\put(05,14){\makebox(0,8)[bl]{\(Z_{5}=Z_{2}\&Z_{3}(2)\)}}
\put(05,10){\makebox(0,8)[bl]{\(Z_{6}=Z_{1}\&Z_{3}(1)\)}}
\put(05,06){\makebox(0,8)[bl]{\(Z_{7}=Z_{1}\&Z_{2}\&Z_{3}(3)\)}}

\put(25,40){\line(0,1){4}}

\put(25,39){\circle*{2}}

\put(27,40){\makebox(0,8)[bl]{\(M=U\star V\)}}

\put(27,36){\makebox(0,8)[bl]{\(M_{1}=U_{1}\star V_{1}\)}}

\put(27,32){\makebox(0,8)[bl]{\(M_{2}=U_{0}\star V_{0}\)}}

\put(25,31){\line(0,1){7}}

\put(25,31){\line(1,0){20}}

\put(35,27){\line(0,1){4}}

\put(35,26){\circle*{2}}

\put(37,26){\makebox(0,8)[bl]{\(U\)}}

\put(33,21){\makebox(0,8)[bl]{\(U_{0}(2)\)}}
\put(33,17){\makebox(0,8)[bl]{\(U_{1}(1)\)}}
\put(33,13){\makebox(0,8)[bl]{\(U_{2}(3)\)}}


\put(45,27){\line(0,1){4}}

\put(45,26){\circle*{2}}

\put(47,26){\makebox(0,8)[bl]{\(V\)}}

\put(43,21){\makebox(0,8)[bl]{\(V_{0}(2)\)}}
\put(43,17){\makebox(0,8)[bl]{\(V_{1}(1)\)}}


\put(55,89){\line(0,1){4}}

\put(56.5,88){\makebox(0,8)[bl]{\(C=H\star Q\)}}

\put(56.5,84){\makebox(0,8)[bl]{\(C_{1}=H_{1}\star Q_{1}\)}}

\put(56.5,80){\makebox(0,8)[bl]{\(C_{2}=H_{1}\star Q_{2}\)}}

\put(55,88){\circle*{2}}

\put(55,74){\line(0,1){13}}

\put(28,74){\line(1,0){27}}


\put(28,70){\line(0,1){4}}

\put(28,69){\circle*{2}}

\put(29.5,69){\makebox(0,8)[bl]{\(H=Y\star G\)}}

\put(29.5,65){\makebox(0,8)[bl]{\(H_{1}=Y_{1}\star G_{1}\)}}

\put(28,64){\line(0,1){4}}

\put(28,64){\line(1,0){10}}

\put(28,60){\line(0,1){4}}

\put(28,59){\circle*{2}}

\put(30,59){\makebox(0,8)[bl]{\(Y\)}}

\put(28,54){\makebox(0,8)[bl]{\(Y_{0}(2)\)}}
\put(28,50){\makebox(0,8)[bl]{\(Y_{1}(1)\)}}
\put(28,46){\makebox(0,8)[bl]{\(Y_{2}(3)\)}}
%


\put(38,60){\line(0,1){4}}

\put(38,59){\circle*{2}}

\put(40,59){\makebox(0,8)[bl]{\(G\)}}

\put(38,54){\makebox(0,8)[bl]{\(G_{0}(2)\)}}
\put(38,50){\makebox(0,8)[bl]{\(G_{1}(1)\)}}


\put(55,70){\line(0,1){4}}

\put(55,69){\circle*{2}}

\put(57,73){\makebox(0,8)[bl]{\(Q=O\star L\)}}

\put(57,69){\makebox(0,8)[bl]{\(Q_{1}=O_{1}\star L_{1}\)}}

\put(57,65){\makebox(0,8)[bl]{\(Q_{2}=O_{1}\star L_{2}\)}}

\put(55,64){\line(0,1){4}}

\put(55,64){\line(1,0){10}}

\put(55,60){\line(0,1){4}}

\put(55,59){\circle*{2}}

\put(57,59){\makebox(0,8)[bl]{\(O\)}}

\put(54,54){\makebox(0,8)[bl]{\(O_{0}(2)\)}}
\put(54,50){\makebox(0,8)[bl]{\(O_{1}(1)\)}}
%


\put(65,60){\line(0,1){4}}

\put(65,59){\circle*{2}}

\put(67,59){\makebox(0,8)[bl]{\(L\)}}

\put(65,54){\makebox(0,8)[bl]{\(L_{0}(2)\)}}
\put(65,50){\makebox(0,8)[bl]{\(L_{1}(1)\)}}
\put(65,46){\makebox(0,8)[bl]{\(L_{2}(1)\)}}

\end{picture}
\end{center}

 {\bf 0.} Plan ~ \( S = A \star B \star C \)

 {\bf 1.} Payment ~ \( A = X \star F \)

 {\it 1.1.} payment scheme ~\( X \):
  100 \% payment \(X_{0}\),
  prepayment of 50...80 percent for parts \(X_{1}\);
  bank loan \(X_{2}\);

 {\it 1.2.} version ~\( F \):
  cash \(F_{1}\), credit card \(F_{2}\),
  bank transfer \(F_{3}\).

 {\bf 2.} Body ~ \( B = R \star Z \star M \):

 {\it 2.1.} frame ~\( W \):
  None \(W_{0}\), technical diagnostics \(W_{1}\),
 follow-up assembly \(W_{2}\);

 {\it 2.2.} hardware ~\( Z \):
  None \(Z_{0}\),
  replacement of defect parts \(Z_{1}\),
  repair of body-defects \(Z_{2}\),
  fitting \(Z_{3}\),
  \(Z_{4}=Z_{1} \& Z_{2} \), \(Z_{5} = Z_{1} \& Z_{3}\),
  \(Z_{6} = Z_{2} \& Z_{3}\), \(Z_{7} = Z_{1} \& Z_{2} \& Z_{3}\);

  {\it 2.3.} finishing ~\( M = U \star  V\):

  {\it 2.3.1.} painting ~\( U \):
   None \(U_{0}\), partial painting \(U_{1}\), painting \(U_{2}\);

  {\it 2.3.2.} appearance restoration ~\( V \):
   None \(V_{0}\), Yes \(V_{1}\).

 {\bf 3.} Electric \& electronic subsystem ~ \( C = H \star  Q \):

 {\it 3.1.} Computer \& navigation subsystem ~ \( H = Y \star  G \):

 {\it 3.1.1.} Computer ~\( Y \):
  None \(Y_{0}\), upgrade \(Y_{1}\),
  additional or new computer \(Y_{2}\);

 {\it 3.1.2.} system GPS ~\( G \):
  None \(G_{0}\), GPS system \(G_{1}\);

 {\it 3.2.} wiring \& lighting  ~ \( Q = O \star  L \):

 {\it 3.2.1.} wiring ~\( O \):
  None \(O_{0}\), repair \(O_{1}\);

 {\it 3.2.2.} lighting  ~\( L \):
  None \(L_{0}\), partial replacement \(L_{1}\),
  replacement \(L_{2}\).

 Tables 12 and 13 contain estimates of compatibility
 (expert judgment).

\begin{center}
\begin{picture}(37,53)

\put(15,49){\makebox(0,0)[bl]{Table 12. Compatibility}}

\put(00,0){\line(1,0){37}} \put(00,42){\line(1,0){37}}
\put(00,48){\line(1,0){37}}

\put(00,0){\line(0,1){48}} \put(07,0){\line(0,1){48}}
\put(37,0){\line(0,1){48}}

\put(01,38){\makebox(0,0)[bl]{\(Z_{0}\)}}
\put(01,34){\makebox(0,0)[bl]{\(Z_{1}\)}}
\put(01,30){\makebox(0,0)[bl]{\(Z_{2}\)}}
\put(01,26){\makebox(0,0)[bl]{\(Z_{3}\)}}
\put(01,22){\makebox(0,0)[bl]{\(Z_{4}\)}}
\put(01,18){\makebox(0,0)[bl]{\(Z_{5}\)}}
\put(01,14){\makebox(0,0)[bl]{\(Z_{6}\)}}
\put(01,10){\makebox(0,0)[bl]{\(Z_{7}\)}}
\put(01,06){\makebox(0,0)[bl]{\(M_{1}\)}}
\put(01,02){\makebox(0,0)[bl]{\(M_{2}\)}}

\put(13,42){\line(0,1){6}} \put(19,42){\line(0,1){6}}
\put(25,42){\line(0,1){6}} \put(31,42){\line(0,1){6}}

\put(07.5,44){\makebox(0,0)[bl]{\(M_{1}\)}}
\put(13.5,44){\makebox(0,0)[bl]{\(M_{2}\)}}
\put(19.5,44){\makebox(0,0)[bl]{\(W_{0}\)}}
\put(25.5,44){\makebox(0,0)[bl]{\(W_{1}\)}}
\put(31.5,44){\makebox(0,0)[bl]{\(W_{2}\)}}

\put(09,38){\makebox(0,0)[bl]{\(2\)}}
\put(15,38){\makebox(0,0)[bl]{\(3\)}}
\put(21,38){\makebox(0,0)[bl]{\(3\)}}
\put(27,38){\makebox(0,0)[bl]{\(3\)}}
\put(33,38){\makebox(0,0)[bl]{\(0\)}}

\put(09,34){\makebox(0,0)[bl]{\(3\)}}
\put(15,34){\makebox(0,0)[bl]{\(2\)}}
\put(21,34){\makebox(0,0)[bl]{\(2\)}}
\put(27,34){\makebox(0,0)[bl]{\(3\)}}
\put(33,34){\makebox(0,0)[bl]{\(3\)}}

\put(09,30){\makebox(0,0)[bl]{\(3\)}}
\put(15,30){\makebox(0,0)[bl]{\(2\)}}
\put(21,30){\makebox(0,0)[bl]{\(0\)}}
\put(27,30){\makebox(0,0)[bl]{\(3\)}}
\put(33,30){\makebox(0,0)[bl]{\(3\)}}

\put(09,26){\makebox(0,0)[bl]{\(3\)}}
\put(15,26){\makebox(0,0)[bl]{\(2\)}}
\put(21,26){\makebox(0,0)[bl]{\(0\)}}
\put(27,26){\makebox(0,0)[bl]{\(2\)}}
\put(33,26){\makebox(0,0)[bl]{\(3\)}}

\put(09,22){\makebox(0,0)[bl]{\(3\)}}
\put(15,22){\makebox(0,0)[bl]{\(2\)}}
\put(21,22){\makebox(0,0)[bl]{\(2\)}}
\put(27,22){\makebox(0,0)[bl]{\(3\)}}
\put(33,22){\makebox(0,0)[bl]{\(3\)}}

\put(09,18){\makebox(0,0)[bl]{\(3\)}}
\put(15,18){\makebox(0,0)[bl]{\(2\)}}
\put(21,18){\makebox(0,0)[bl]{\(0\)}}
\put(27,18){\makebox(0,0)[bl]{\(3\)}}
\put(33,18){\makebox(0,0)[bl]{\(3\)}}

\put(09,14){\makebox(0,0)[bl]{\(3\)}}
\put(15,14){\makebox(0,0)[bl]{\(2\)}}
\put(21,14){\makebox(0,0)[bl]{\(2\)}}
\put(27,14){\makebox(0,0)[bl]{\(3\)}}
\put(33,14){\makebox(0,0)[bl]{\(3\)}}

\put(09,10){\makebox(0,0)[bl]{\(3\)}}
\put(15,10){\makebox(0,0)[bl]{\(2\)}}
\put(21,10){\makebox(0,0)[bl]{\(2\)}}
\put(27,10){\makebox(0,0)[bl]{\(3\)}}
\put(33,10){\makebox(0,0)[bl]{\(3\)}}

\put(21,06){\makebox(0,0)[bl]{\(0\)}}
\put(27,06){\makebox(0,0)[bl]{\(3\)}}
\put(33,06){\makebox(0,0)[bl]{\(3\)}}

\put(21,02){\makebox(0,0)[bl]{\(3\)}}
\put(27,02){\makebox(0,0)[bl]{\(2\)}}
\put(33,02){\makebox(0,0)[bl]{\(2\)}}

\end{picture}
%
\begin{picture}(25,27)

\put(00,0){\line(1,0){25}} \put(00,14){\line(1,0){25}}
\put(00,20){\line(1,0){25}}

\put(00,0){\line(0,1){20}} \put(07,0){\line(0,1){20}}
\put(25,0){\line(0,1){20}}

\put(01,10){\makebox(0,0)[bl]{\(X_{1}\)}}
\put(01,06){\makebox(0,0)[bl]{\(X_{2}\)}}
\put(01,02){\makebox(0,0)[bl]{\(X_{3}\)}}

\put(13,14){\line(0,1){6}} \put(19,14){\line(0,1){6}}

\put(07.5,16){\makebox(0,0)[bl]{\(F_{1}\)}}
\put(13.5,16){\makebox(0,0)[bl]{\(F_{2}\)}}
\put(19.5,16){\makebox(0,0)[bl]{\(F_{3}\)}}

\put(09,10){\makebox(0,0)[bl]{\(3\)}}
\put(15,10){\makebox(0,0)[bl]{\(3\)}}
\put(21,10){\makebox(0,0)[bl]{\(3\)}}

\put(09,06){\makebox(0,0)[bl]{\(3\)}}
\put(15,06){\makebox(0,0)[bl]{\(3\)}}
\put(21,06){\makebox(0,0)[bl]{\(3\)}}

\put(09,02){\makebox(0,0)[bl]{\(0\)}}
\put(15,02){\makebox(0,0)[bl]{\(3\)}}
\put(21,02){\makebox(0,0)[bl]{\(2\)}}


\put(00,32){\line(1,0){25}} \put(00,42){\line(1,0){25}}
\put(00,48){\line(1,0){25}}

\put(00,32){\line(0,1){16}} \put(07,32){\line(0,1){16}}
\put(25,32){\line(0,1){16}}

\put(01,38){\makebox(0,0)[bl]{\(O_{0}\)}}
\put(01,34){\makebox(0,0)[bl]{\(O_{1}\)}}

\put(13,42){\line(0,1){6}} \put(19,42){\line(0,1){6}}

\put(07.5,44){\makebox(0,0)[bl]{\(L_{0}\)}}
\put(13.5,44){\makebox(0,0)[bl]{\(L_{1}\)}}
\put(19.5,44){\makebox(0,0)[bl]{\(L_{2}\)}}

\put(09,38){\makebox(0,0)[bl]{\(3\)}}
\put(15,38){\makebox(0,0)[bl]{\(2\)}}
\put(21,38){\makebox(0,0)[bl]{\(2\)}}

\put(09,34){\makebox(0,0)[bl]{\(1\)}}
\put(15,34){\makebox(0,0)[bl]{\(3\)}}
\put(21,34){\makebox(0,0)[bl]{\(3\)}}


\end{picture}
\end{center}

\begin{center}
\begin{picture}(19,25)

\put(03,21){\makebox(0,0)[bl]{Table 13. Compatibility}}

\put(00,0){\line(1,0){19}} \put(00,14){\line(1,0){19}}
\put(00,20){\line(1,0){19}}

\put(00,0){\line(0,1){20}} \put(07,0){\line(0,1){20}}
\put(19,0){\line(0,1){20}}

\put(01,10){\makebox(0,0)[bl]{\(Y_{0}\)}}
\put(01,06){\makebox(0,0)[bl]{\(Y_{1}\)}}
\put(01,02){\makebox(0,0)[bl]{\(Y_{2}\)}}

\put(13,14){\line(0,1){6}}

\put(07.5,16){\makebox(0,0)[bl]{\(G_{0}\)}}
\put(13.5,16){\makebox(0,0)[bl]{\(G_{1}\)}}

\put(09,10){\makebox(0,0)[bl]{\(3\)}}
\put(15,10){\makebox(0,0)[bl]{\(0\)}}

\put(09,06){\makebox(0,0)[bl]{\(2\)}}
\put(15,06){\makebox(0,0)[bl]{\(3\)}}

\put(09,02){\makebox(0,0)[bl]{\(1\)}}
\put(15,02){\makebox(0,0)[bl]{\(2\)}}

\end{picture}
%
\begin{picture}(19,28)

\put(00,0){\line(1,0){19}} \put(00,14){\line(1,0){19}}
\put(00,20){\line(1,0){19}}

\put(00,0){\line(0,1){20}} \put(07,0){\line(0,1){20}}
\put(19,0){\line(0,1){20}}

\put(01,10){\makebox(0,0)[bl]{\(U_{0}\)}}
\put(01,06){\makebox(0,0)[bl]{\(U_{1}\)}}
\put(01,02){\makebox(0,0)[bl]{\(U_{2}\)}}

\put(13,14){\line(0,1){6}}

\put(07.5,16){\makebox(0,0)[bl]{\(V_{0}\)}}
\put(13.5,16){\makebox(0,0)[bl]{\(V_{1}\)}}

\put(09,10){\makebox(0,0)[bl]{\(3\)}}
\put(15,10){\makebox(0,0)[bl]{\(0\)}}

\put(09,06){\makebox(0,0)[bl]{\(0\)}}
\put(15,06){\makebox(0,0)[bl]{\(2\)}}

\put(09,02){\makebox(0,0)[bl]{\(0\)}}
\put(15,02){\makebox(0,0)[bl]{\(3\)}}

\end{picture}
\end{center}

 The following intermediate composite Pareto-efficient DAs are obtained:

 \(A_{1} = X_{1} \star F_{2}\), \(N(A_{1}) = (3;2,0,0)\);

 \(H_{1} = Y_{1} \star G_{1}\), \(N(H_{1}) = (3;2,0,0)\);

 \(Q_{1} = O_{1} \star L_{1}\), \(N(Q_{1}) = (3;2,0,0)\);
 \(Q_{2} = O_{1} \star L_{2}\), \(N(Q_{2}) = (3;2,0,0)\);

 \(M_{1} = U_{1} \star V_{2}\), \(N(M_{1}) = (2;2,0,0)\);
 \(M_{2} = U_{0} \star V_{0}\), \(N(M_{2}) = (3;0,2,0)\);

 \(B_{1} = W_{1} \star Z_{1} \star M_{1}\), \(N(B_{1}) = (3;3,0,0)\);
 \(B_{2} = W_{1} \star Z_{6} \star M_{1}\), \(N(B_{2}) =
 (3;3,0,0)\).

 The resultant composite
 Pareto-efficient
 DAs are the  following
 (for a final user's analysis/choice):
 \(S_{1}=A_{1}\star B_{1} \star C_{1}\),
 \(S_{2}=A_{1}\star B_{2} \star C_{1}\),
 \(S_{3}=A_{1}\star B_{1} \star C_{2}\), and
 \(S_{4}=A_{1}\star B_{2} \star C_{2}\).

\subsection{Synthesis of Product Trajectory}

 In addition,
  it is reasonable to consider the design problem for synthesis of product (system)
 trajectory
%
  as follows
 (e.g., \cite{lev06}):


  {\it  Combine a trajectory
  (i.e., selection
 of a system solution
  at each time stage )
  while taking into account quality of composite DAs at each time stage
  and a cost of the component changes.}

 Let us consider a three-stage example:
 (i) computer for stage 1 (Fig. 4),
 (ii) computer for stage 1 (Fig. 15), and
 (iii) computer for stage 1 (Fig. 16).
 Here the tree-like structure and DAs are the same,
 priorities of DAs are different
 (priorities are shown in parentheses in Fig. 4, Fig. 15, Fig. 16),
 estimates of compatibility between DAs are the same (Table 3)
 (estimates have an illustrative character and are based on expert judgment).


\begin{center}
\begin{picture}(78,76)
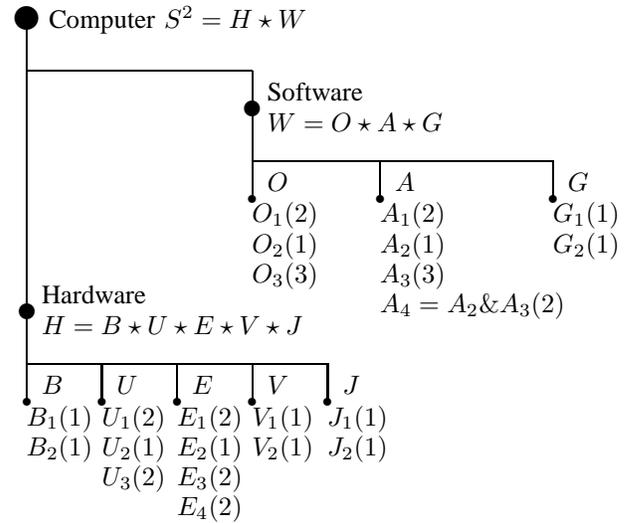
\put(10,0){\makebox(0,0)[bl] {Fig. 15. Personal computer (stage
2)}}


\put(00,72){\circle*{3}}

\put(03,70){\makebox(0,0)[bl]{Computer \(S^{2} = H \star W  \) }}

\put(00,64){\line(0,1){8}}



\put(30,65){\line(-1,0){30}}

\put(32,61){\makebox(0,0)[bl]{Software}}

\put(32,57){\makebox(0,0)[bl]{\(W = O \star A \star G\)}}


\put(30,60){\circle*{2}} \put(30,57){\line(0,1){8}}

\put(30,53){\line(1,0){40}}


\put(30,48){\line(0,1){10}}


\put(47,48){\line(0,1){05}} \put(70,48){\line(0,1){05}}

\put(30,48){\circle*{1}}


\put(47,48){\circle*{1}} \put(70,48){\circle*{1}}


\put(32,49){\makebox(0,0)[bl]{\(O\)}}
\put(49,49){\makebox(0,0)[bl]{\(A\)}}
\put(72,49){\makebox(0,0)[bl]{\(G\)}}

\put(70,44){\makebox(0,0)[bl]{\(G_{1}(1)\)}}
\put(70,40){\makebox(0,0)[bl]{\(G_{2}(1)\)}}

\put(47,44){\makebox(0,0)[bl]{\(A_{1}(2)\)}}
\put(47,40){\makebox(0,0)[bl]{\(A_{2}(1)\)}}
\put(47,36){\makebox(0,0)[bl]{\(A_{3}(3)\)}}
\put(47,32){\makebox(0,0)[bl]{\(A_{4}=A_{2}\&A_{3}(2)\)}}


\put(30,44){\makebox(0,0)[bl]{\(O_{1}(2)\)}}
\put(30,40){\makebox(0,0)[bl]{\(O_{2}(1)\)}}
\put(30,36){\makebox(0,0)[bl]{\(O_{3}(3)\)}}


\put(02,34){\makebox(0,0)[bl]{Hardware}}

\put(02,30){\makebox(0,0)[bl]{\(H = B \star U \star E \star V
\star J\)}}

\put(00,42){\line(0,1){30}}


\put(00,37){\line(0,1){05}} \put(00,33){\circle*{2}}
\put(00,30){\line(0,1){11}}

\put(00,26){\line(1,0){40}}


\put(00,21){\line(0,1){10}} \put(10,21){\line(0,1){05}}
\put(20,21){\line(0,1){05}} \put(30,21){\line(0,1){05}}
\put(40,21){\line(0,1){05}}

\put(00,21){\circle*{1}} \put(10,21){\circle*{1}}
\put(20,21){\circle*{1}} \put(30,21){\circle*{1}}
\put(40,21){\circle*{1}}


\put(02,22){\makebox(0,0)[bl]{\(B\)}}
\put(12,22){\makebox(0,0)[bl]{\(U\)}}
\put(22,22){\makebox(0,0)[bl]{\(E\)}}
\put(32,22){\makebox(0,0)[bl]{\(V\)}}
\put(42,22){\makebox(0,0)[bl]{\(J\)}}

\put(40,17){\makebox(0,0)[bl]{\(J_{1}(1)\)}}
\put(40,13){\makebox(0,0)[bl]{\(J_{2}(1)\)}}

\put(30,17){\makebox(0,0)[bl]{\(V_{1}(1)\)}}
\put(30,13){\makebox(0,0)[bl]{\(V_{2}(1)\)}}

\put(20,17){\makebox(0,0)[bl]{\(E_{1}(2)\)}}
\put(20,13){\makebox(0,0)[bl]{\(E_{2}(1)\)}}
\put(20,09){\makebox(0,0)[bl]{\(E_{3}(2)\)}}
\put(20,05){\makebox(0,0)[bl]{\(E_{4}(2)\)}}

\put(10,17){\makebox(0,0)[bl]{\(U_{1}(2)\)}}
\put(10,13){\makebox(0,0)[bl]{\(U_{2}(1)\)}}
\put(10,09){\makebox(0,0)[bl]{\(U_{3}(2)\)}}

\put(00,17){\makebox(0,0)[bl]{\(B_{1}(1)\)}}
\put(00,13){\makebox(0,0)[bl]{\(B_{2}(1)\)}}

\end{picture}
\end{center}


\begin{center}
\begin{picture}(78,76)
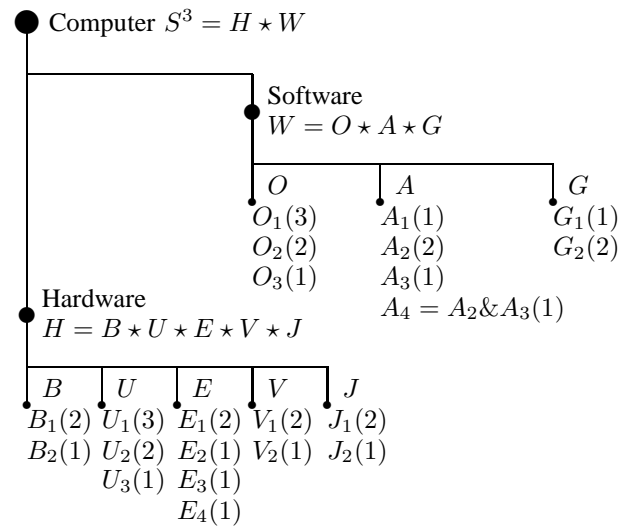
\put(10,0){\makebox(0,0)[bl] {Fig. 16. Personal computer (stage
3)}}


\put(00,72){\circle*{3}}

\put(03,70){\makebox(0,0)[bl]{Computer \(S^{3} = H \star W  \) }}

\put(00,64){\line(0,1){8}}



\put(30,65){\line(-1,0){30}}

\put(32,61){\makebox(0,0)[bl]{Software}}

\put(32,57){\makebox(0,0)[bl]{\(W = O \star A \star G\)}}


\put(30,60){\circle*{2}} \put(30,57){\line(0,1){8}}

\put(30,53){\line(1,0){40}}


\put(30,48){\line(0,1){10}}


\put(47,48){\line(0,1){05}} \put(70,48){\line(0,1){05}}

\put(30,48){\circle*{1}}


\put(47,48){\circle*{1}} \put(70,48){\circle*{1}}


\put(32,49){\makebox(0,0)[bl]{\(O\)}}
\put(49,49){\makebox(0,0)[bl]{\(A\)}}
\put(72,49){\makebox(0,0)[bl]{\(G\)}}

\put(70,44){\makebox(0,0)[bl]{\(G_{1}(1)\)}}
\put(70,40){\makebox(0,0)[bl]{\(G_{2}(2)\)}}

\put(47,44){\makebox(0,0)[bl]{\(A_{1}(1)\)}}
\put(47,40){\makebox(0,0)[bl]{\(A_{2}(2)\)}}
\put(47,36){\makebox(0,0)[bl]{\(A_{3}(1)\)}}
\put(47,32){\makebox(0,0)[bl]{\(A_{4}=A_{2}\&A_{3}(1)\)}}


\put(30,44){\makebox(0,0)[bl]{\(O_{1}(3)\)}}
\put(30,40){\makebox(0,0)[bl]{\(O_{2}(2)\)}}
\put(30,36){\makebox(0,0)[bl]{\(O_{3}(1)\)}}


\put(02,34){\makebox(0,0)[bl]{Hardware}}

\put(02,30){\makebox(0,0)[bl]{\(H = B \star U \star E \star V
\star J\)}}

\put(00,42){\line(0,1){30}}


\put(00,37){\line(0,1){05}} \put(00,33){\circle*{2}}
\put(00,30){\line(0,1){11}}

\put(00,26){\line(1,0){40}}


\put(00,21){\line(0,1){10}} \put(10,21){\line(0,1){05}}
\put(20,21){\line(0,1){05}} \put(30,21){\line(0,1){05}}
\put(40,21){\line(0,1){05}}

\put(00,21){\circle*{1}} \put(10,21){\circle*{1}}
\put(20,21){\circle*{1}} \put(30,21){\circle*{1}}
\put(40,21){\circle*{1}}


\put(02,22){\makebox(0,0)[bl]{\(B\)}}
\put(12,22){\makebox(0,0)[bl]{\(U\)}}
\put(22,22){\makebox(0,0)[bl]{\(E\)}}
\put(32,22){\makebox(0,0)[bl]{\(V\)}}
\put(42,22){\makebox(0,0)[bl]{\(J\)}}

\put(40,17){\makebox(0,0)[bl]{\(J_{1}(2)\)}}
\put(40,13){\makebox(0,0)[bl]{\(J_{2}(1)\)}}

\put(30,17){\makebox(0,0)[bl]{\(V_{1}(2)\)}}
\put(30,13){\makebox(0,0)[bl]{\(V_{2}(1)\)}}

\put(20,17){\makebox(0,0)[bl]{\(E_{1}(2)\)}}
\put(20,13){\makebox(0,0)[bl]{\(E_{2}(1)\)}}
\put(20,09){\makebox(0,0)[bl]{\(E_{3}(1)\)}}
\put(20,05){\makebox(0,0)[bl]{\(E_{4}(1)\)}}

\put(10,17){\makebox(0,0)[bl]{\(U_{1}(3)\)}}
\put(10,13){\makebox(0,0)[bl]{\(U_{2}(2)\)}}
\put(10,09){\makebox(0,0)[bl]{\(U_{3}(1)\)}}

\put(00,17){\makebox(0,0)[bl]{\(B_{1}(2)\)}}
\put(00,13){\makebox(0,0)[bl]{\(B_{2}(1)\)}}

\end{picture}
\end{center}

 The  following composite solutions are obtained (Fig. 4, Fig. 15, Fig. 16):

 {\it Stage 1}:

 \(H_{1} = B_{1} \star U_{1}\star E_{1} \star V_{1} \star J_{1} \),
 \(N(H_{1})=(3;5,0,0)\),

 \(W_{1} = O_{1} \star A_{1}\star G_{2} \),
 \(N(W_{1})=(2;3,0,0)\),

 \(W_{2} = O_{2} \star A_{1}\star G_{2} \),
 \(N(W_{2})=(3;2,1,0)\);

  \(S_{1} = H_{1} \star W_{1}= (B_{1} \star U_{1}\star E_{1} \star V_{1} \star J_{1}) \star (O_{1} \star A_{1}\star G_{2})\),

  \(S_{2} = H_{1} \star W_{2}= (B_{1} \star U_{1}\star E_{1} \star V_{1} \star J_{1}) \star (O_{2} \star A_{1}\star G_{2}) \).

 {\it Stage 2}:

 \(H^{2}_{1} = B_{2} \star U_{2}\star E_{2} \star V_{2} \star J_{2} \),
 \(N(H^{2}_{1})=(3;5,0,0)\),

 \(W^{2}_{1} = O_{2} \star A_{2}\star G_{1} \),
 \(N(W^{2}_{1})=(3;3,0,0)\),

 \(W^{2}_{2} = O_{2} \star A_{2}\star G_{2} \),
 \(N(W^{2}_{2})=(3;3,0,0)\);

  \(S^{2}_{1} = H^{2}_{1} \star W^{2}_{1}= (B_{2} \star U_{2}\star E_{2} \star V_{2} \star J_{2}) \star (O_{2} \star A_{2}\star G_{1})\),

  \(S^{2}_{2} = H^{2}_{1} \star W^{2}_{2}= (B_{2} \star U_{2}\star E_{2} \star V_{2} \star J_{2}) \star (O_{2} \star A_{2}\star G_{2}) \).

  {\it Stage 3}:

 \(H^{3}_{1} = B_{2} \star U_{3}\star E_{2} \star V_{2} \star J_{2} \),
 \(N(H^{3}_{1})=(3;5,0,0)\),

 \(W^{3}_{1} = O_{3} \star A_{1}\star G_{1} \),
 \(N(W^{3}_{1})=(2;3,0,0)\),

 \(W^{3}_{2} = O_{3} \star A_{3}\star G_{1} \),
 \(N(W^{3}_{2})=(3;2,1,0)\);

  \(S^{3}_{1} = H^{3}_{1} \star W^{3}_{1}= (B_{2} \star U_{3}\star E_{2} \star V_{2} \star J_{2}) \star (O_{3} \star A_{1}\star G_{1})\),

  \(S^{3}_{2} = H^{3}_{1} \star W^{3}_{2}= (B_{2} \star U_{3}\star E_{2} \star V_{2} \star J_{2}) \star (O_{3} \star A_{3}\star G_{1}) \).

 Table 14 contains the numbers of element changes for products at
 different stages
 (products at neighborhood stages are compared):
 ~\(\delta (S_{1},S^{2}_{1})\), etc.
 The estimate of compatibility is computed as follows (Table 15):
  ~\( \xi (S_{1},S^{2}_{1}) = (8-\delta (S_{1},S^{2}_{1}))\).

\begin{center}
\begin{picture}(43,36)
\put(00,31){\makebox(0,0)[bl]{Table 14.  Changes \(\delta
(S',S'')\)}}

\put(00,0){\line(1,0){35}} \put(00,22){\line(1,0){35}}
\put(00,29){\line(1,0){35}}

\put(00,0){\line(0,1){29}} \put(06.5,0){\line(0,1){29}}
\put(35,0){\line(0,1){29}}

\put(01,17){\makebox(0,0)[bl]{\(S_{1}\)}}
\put(01,12){\makebox(0,0)[bl]{\(S_{2}\)}}
\put(01,07){\makebox(0,0)[bl]{\(S^{2}_{1}\)}}
\put(01,02){\makebox(0,0)[bl]{\(S^{2}_{2}\)}}

\put(14,22){\line(0,1){7}} \put(21,22){\line(0,1){7}}
\put(28,22){\line(0,1){7}}

\put(08.4,23.5){\makebox(0,0)[bl]{\(S^{2}_{1}\)}}
\put(15.4,23.5){\makebox(0,0)[bl]{\(S^{2}_{2}\)}}
\put(22.4,23.5){\makebox(0,0)[bl]{\(S^{3}_{1}\)}}
\put(29.4,23.5){\makebox(0,0)[bl]{\(S^{3}_{2}\)}}


\put(09,17){\makebox(0,0)[bl]{\(8\)}}
\put(16,17){\makebox(0,0)[bl]{\(7\)}}
\put(23,17){\makebox(0,0)[bl]{\(-\)}}
\put(30,17){\makebox(0,0)[bl]{\(-\)}}

\put(09,12){\makebox(0,0)[bl]{\(7\)}}
\put(16,12){\makebox(0,0)[bl]{\(6\)}}
\put(23,12){\makebox(0,0)[bl]{\(-\)}}
\put(30,12){\makebox(0,0)[bl]{\(-\)}}

\put(23,07){\makebox(0,0)[bl]{\(3\)}}
\put(30,07){\makebox(0,0)[bl]{\(3\)}}

\put(23,02){\makebox(0,0)[bl]{\(3\)}}
\put(30,02){\makebox(0,0)[bl]{\(4\)}}



\end{picture}
%
%
\begin{picture}(35,36)
\put(00,31){\makebox(0,0)[bl]{Table 15. Compatibility}}

\put(00,0){\line(1,0){35}} \put(00,22){\line(1,0){35}}
\put(00,29){\line(1,0){35}}

\put(00,0){\line(0,1){29}} \put(06.5,0){\line(0,1){29}}
\put(35,0){\line(0,1){29}}

\put(01,17){\makebox(0,0)[bl]{\(S_{1}\)}}
\put(01,12){\makebox(0,0)[bl]{\(S_{2}\)}}
\put(01,07){\makebox(0,0)[bl]{\(S^{2}_{1}\)}}
\put(01,02){\makebox(0,0)[bl]{\(S^{2}_{2}\)}}

\put(14,22){\line(0,1){7}} \put(21,22){\line(0,1){7}}
\put(28,22){\line(0,1){7}}

\put(08.4,23.5){\makebox(0,0)[bl]{\(S^{2}_{1}\)}}
\put(15.4,23.5){\makebox(0,0)[bl]{\(S^{2}_{2}\)}}
\put(22.4,23.5){\makebox(0,0)[bl]{\(S^{3}_{1}\)}}
\put(29.4,23.5){\makebox(0,0)[bl]{\(S^{3}_{2}\)}}


\put(09,17){\makebox(0,0)[bl]{\(0\)}}
\put(16,17){\makebox(0,0)[bl]{\(1\)}}
\put(23,17){\makebox(0,0)[bl]{\(-\)}}
\put(30,17){\makebox(0,0)[bl]{\(-\)}}

\put(09,12){\makebox(0,0)[bl]{\(1\)}}
\put(16,12){\makebox(0,0)[bl]{\(2\)}}
\put(23,12){\makebox(0,0)[bl]{\(-\)}}
\put(30,12){\makebox(0,0)[bl]{\(-\)}}

\put(23,07){\makebox(0,0)[bl]{\(5\)}}
\put(30,07){\makebox(0,0)[bl]{\(5\)}}

\put(23,02){\makebox(0,0)[bl]{\(5\)}}
\put(30,02){\makebox(0,0)[bl]{\(4\)}}



\end{picture}
\end{center}

 The designed trajectory is based on combinatorial synthesis.
 It is assumed, the composite solutions for stages 1, 2, and 3
 (i.e., \(S_{1}\), \(S_{2}\), \(S^{2}_{1}\),
 \(S^{2}_{2}\), \(S^{3}_{1}\), \(S^{3}_{2}\))
 have priorities at the level \(1\).
 The best composition
 (while taking into account compatibility estimates)
 is (Fig. 17):
  ~\(\overline{\alpha} = < S_{2}, S^{2}_{2}, S^{3}_{1} > \),

\begin{center}
\begin{picture}(80,22)
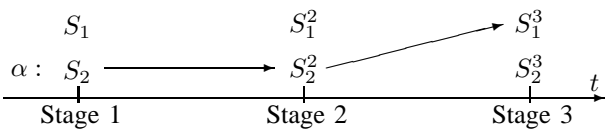
\put(13,0){\makebox(0,0)[bl] {Fig. 17. Design of
 system trajectory}}

\put(0,8){\vector(1,0){80}} \put(78,9){\makebox(0,8)[bl]{\(t\)}}

\put(5,4){\makebox(0,8)[bl]{Stage 1}}
\put(35,4){\makebox(0,8)[bl]{Stage 2}}
\put(65,4){\makebox(0,8)[bl]{Stage 3}}

\put(10,7.5){\line(0,1){2}} \put(40,7.5){\line(0,1){2}}
\put(70,7.5){\line(0,1){2}}

\put(01,11){\makebox(0,8)[bl]{\(\alpha :\)}}

\put(13.5,12){\vector(1,0){22.6}}



\put(43,12){\vector(4,1){23.5}}


\put(8,16){\makebox(0,8)[bl]{\(S_{1}\)}}

\put(8,10){\makebox(0,8)[bl]{\(S_{2}\)}}



\put(38,16){\makebox(0,8)[bl]{\(S^{2}_{1}\)}}


\put(38,10){\makebox(0,8)[bl]{\(S^{2}_{2}\)}}


\put(68,16){\makebox(0,8)[bl]{\(S^{3}_{1}\)}}

\put(68,10){\makebox(0,8)[bl]{\(S^{3}_{2}\)}}



\end{picture}
\end{center}

\subsection{Aggregation of Modular Products}
%

 An example of aggregation process is based on
%
 the multi-choice scheme
 with aggregation (Fig. 7).
 An initial morphological structure
 of a car is the following  (Fig. 18)
 (in real application, this structure can be considered
 as a result of processing the selected products/solutions)
 \cite{lev11agg}:

 {\bf 0.} Car
 ~\(S = A \star B \star C \).

 {\bf 1.} Main part
 \(A = E \star D \):

 {\it 1.1.} Engine  E:~
 diesel \(E_{1}\),
 gasoline \(E_{2}\),
 electric \(E_{3}\),
 hydrogenous \(E_{4}\), and
 hybrid synergy drive HSD \(E_{5}\);

 {\it 1.2.} Body \(D\):~
 sedan \(D_{1}\),
 universal \(D_{2}\),
 jeep \(D_{3}\),
 pickup \(D_{4}\), and
 sport \(D_{5}\).

 {\bf 2.} Mechanical part
   \(B = X \star Y \star  Z \):

  {\it 2.1.} gear box
    X:~
  automate  \(X_{1}\),
  manual \(X_{2}\);

 {\it 2.2.} suspension
 Y:~
 pneumatic  \(Y_{1}\),
 hydraulic  \(Y_{2}\), and
 pneumohydraulic \(Y_{3}\);

 {\it 2.3.} drive
 Z:~
 front-wheel drive  \(Z_{1}\),
 rear-drive \(Z_{2}\),
 all-wheel-drive \(Z_{3}\).

 {\bf 3.} Safety part
 \(C = O \star G  \):

  {\it 3.1.}
  \(O\):~
  ``absence'' \(O_{0}\),
 electronic \(O_{1}\);

 {\it 3.2.} Safety subsystem
 \(G\):
  ``absence'' \(G_{0}\),
  passive \(G_{1}\),
  active \(G_{2}\).


\begin{center}
\begin{picture}(70,99)

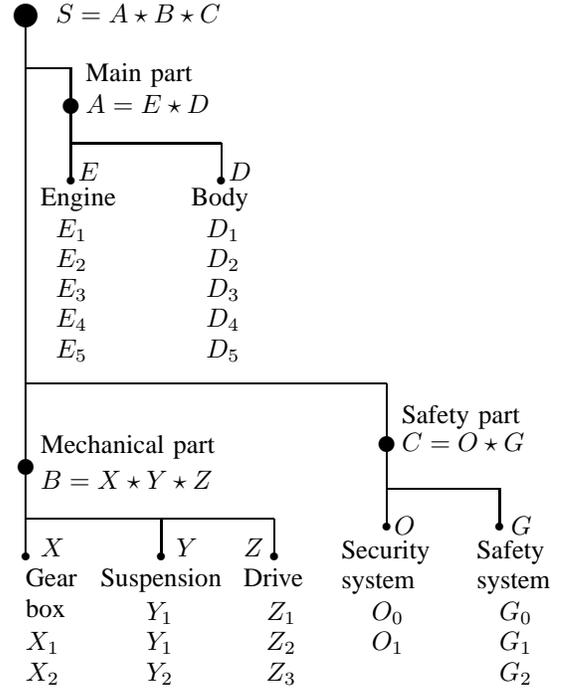
\put(08,00){\makebox(0,0)[bl] {Fig. 18. General structure
 of car
 }}

\put(00,95){\circle*{3}}


\put(04,94){\makebox(0,0)[bl]{\(S = A \star B \star C \) }}

\put(00,88){\line(0,1){7}}


\put(08,85.5){\makebox(0,0)[bl]{Main part}}

\put(08,82){\makebox(0,0)[bl]{\(A = E \star D \)}}

\put(06,83){\line(0,1){5}}

\put(00,88){\line(1,0){6}}


\put(06,83){\line(0,1){05}} \put(06,83){\circle*{2}}

\put(06,78){\line(1,0){20}}


\put(06,73){\line(0,1){10}} \put(26,73){\line(0,1){05}}

\put(06,73){\circle*{1}} \put(26,73){\circle*{1}}

\put(7,73){\makebox(0,0)[bl]{\(E\)}}
\put(27,73){\makebox(0,0)[bl]{\(D\)}}


\put(2,69){\makebox(0,0)[bl]{Engine}}
\put(4,65){\makebox(0,0)[bl]{\(E_{1}\)}}
\put(4,61){\makebox(0,0)[bl]{\(E_{2}\)}}
\put(4,57){\makebox(0,0)[bl]{\(E_{3}\)}}
\put(4,53){\makebox(0,0)[bl]{\(E_{4}\)}}
\put(4,49){\makebox(0,0)[bl]{\(E_{5}\)}}

\put(22,69){\makebox(0,0)[bl]{Body}}
\put(24,65){\makebox(0,0)[bl]{\(D_{1}\)}}
\put(24,61){\makebox(0,0)[bl]{\(D_{2}\)}}
\put(24,57){\makebox(0,0)[bl]{\(D_{3}\)}}
\put(24,53){\makebox(0,0)[bl]{\(D_{4}\)}}
\put(24,49){\makebox(0,0)[bl]{\(D_{5}\)}}


\put(50,40){\makebox(0,0)[bl]{Safety part}}

\put(50,37){\makebox(0,0)[bl]{\(C = O \star G \)}}

\put(48,41){\line(0,1){5}}

\put(00,46){\line(1,0){48}}


\put(48,33){\line(0,1){08}} \put(48,38){\circle*{2}}


\put(48,32){\line(1,0){15}}


\put(48,27){\line(0,1){13}}

\put(63,27){\line(0,1){05}}


\put(48,27){\circle*{1}} \put(63,27){\circle*{1}}


\put(49,26){\makebox(0,0)[bl]{\(O\)}}

\put(64.5,26){\makebox(0,0)[bl]{\(G \)}}

\put(60,22){\makebox(0,0)[bl]{Safety}}
\put(60,18){\makebox(0,0)[bl]{system}}
\put(63,14){\makebox(0,0)[bl]{\(G_{0}\)}}
\put(63,10){\makebox(0,0)[bl]{\(G_{1}\)}}
\put(63,06){\makebox(0,0)[bl]{\(G_{2}\)}}


\put(42,22){\makebox(0,0)[bl]{Security}}
\put(42,18){\makebox(0,0)[bl]{system}}
\put(46,14){\makebox(0,0)[bl]{\(O_{0}\)}}
\put(46,10){\makebox(0,0)[bl]{\(O_{1}\)}}


\put(02,36){\makebox(0,0)[bl]{Mechanical part}}

\put(02,32){\makebox(0,0)[bl]{\(B = X \star Y \star Z \)}}

\put(00,44){\line(0,1){50}}


\put(00,39){\line(0,1){05}} \put(00,35){\circle*{2}}
\put(00,32){\line(0,1){11}}


\put(00,28){\line(1,0){33}}


\put(00,23){\line(0,1){10}} \put(18,23){\line(0,1){05}}
\put(33,23){\line(0,1){05}}


\put(00,23){\circle*{1}} \put(18,23){\circle*{1}}
\put(33,23){\circle*{1}}



\put(02,23){\makebox(0,0)[bl]{\(X\)}}
\put(20,23){\makebox(0,0)[bl]{\(Y\)}}
\put(29,23){\makebox(0,0)[bl]{\(Z\)}}


\put(29,19){\makebox(0,0)[bl]{Drive}}

\put(32,14){\makebox(0,0)[bl]{\(Z_{1}\)}}
\put(32,10){\makebox(0,0)[bl]{\(Z_{2}\)}}
\put(32,6){\makebox(0,0)[bl]{\(Z_{3}\)}}

\put(10,18.3){\makebox(0,0)[bl]{Suspension}}
\put(16,14){\makebox(0,0)[bl]{\(Y_{1}\)}}
\put(16,10){\makebox(0,0)[bl]{\(Y_{1}\)}}
\put(16,06){\makebox(0,0)[bl]{\(Y_{2}\)}}

\put(00,19){\makebox(0,0)[bl]{Gear}}
\put(00,15){\makebox(0,0)[bl]{box}}
\put(00,10){\makebox(0,0)[bl]{\(X_{1}\)}}
\put(00,06){\makebox(0,0)[bl]{\(X_{2}\)}}

\end{picture}
\end{center}

 The following initial solutions/prototypes
  are considered \cite{lev11agg}:

 \(S^{1}_{1}=E_{1}\star D_{1}\star X_{1}\star Y_{1}\star Z_{1}
 \star O_{1}\star G_{1} \),

 \(S^{1}_{2}=E_{5}\star D_{1}\star X_{1}\star Y_{1}\star Z_{1}
 \star O_{1}\star G_{2} \),

  \(S^{2}_{1}=E_{2}\star D_{1}\star X_{2}\star Y_{1}\star Z_{1}
 \star O_{0}\star G_{1} \),

  \(S^{3}_{1}=E_{2}\star D_{3}\star X_{1}\star Y_{2}\star Z_{3}
 \star O_{1}\star G_{0} \), and

  \(S^{3}_{2}=E_{2}\star D_{5}\star X_{1}\star Y_{3}\star Z_{1}
 \star O_{1}\star G_{1} \).

 The substructure of the five solutions above is empty.
 A ``kernel'' can be designed
 by the following element inclusion rule:

 {\it component} ~\(\iota\)~
 {\it is included into the ``kernel'' if}
 ~\( \eta_{\iota}  \geq \lambda \),

 where
 \(\eta_{\iota}\) is the number of DAs \(\iota\) in initial
 prototypes/products,
 \(\lambda \leq m   \),
 \(m\) is the number of initial prototypes/product.
 The obtained ``kernel'' (as a basis for extension)
 is presented in Fig. 19
  (here \(\lambda = 2  \)).
  The superstructure is presented in Fig. 20.

\begin{center}
\begin{picture}(59,22)

\put(11,00){\makebox(0,0)[bl] {Fig. 19. ``System kernel''}}


\put(05,16){\circle*{1.5}} \put(13,16){\circle*{1.5}}
\put(21,16){\circle*{1.5}} \put(29,16){\circle*{1.5}}
\put(37,16){\circle*{1.5}} \put(45,16){\circle*{1.5}}
\put(53,16){\circle*{1.5}}

\put(04,18){\makebox(0,0)[bl]{\(E\)}}
\put(12,18){\makebox(0,0)[bl]{\(D\)}}
\put(20,18){\makebox(0,0)[bl]{\(X\)}}
\put(28,18){\makebox(0,0)[bl]{\(Y\)}}
\put(35,18){\makebox(0,0)[bl]{\(Z\)}}

\put(43,18){\makebox(0,0)[bl]{\(O\)}}
\put(52,18){\makebox(0,0)[bl]{\(G\)}}


\put(05,11){\oval(07,10)} \put(13,11){\oval(07,10)}
\put(21,11){\oval(07,10)} \put(29,11){\oval(07,10)}
\put(37,11){\oval(07,10)} \put(45,11){\oval(07,10)}
\put(53,11){\oval(07,10)}

\put(02.8,11){\makebox(0,0)[bl]{\(E_{2}\)}}

\put(10.8,11){\makebox(0,0)[bl]{\(D_{1}\)}}

\put(18.8,11){\makebox(0,0)[bl]{\(X_{1}\)}}

\put(26.8,11){\makebox(0,0)[bl]{\(Y_{1}\)}}

\put(34.8,11){\makebox(0,0)[bl]{\(Z_{1}\)}}

\put(42.8,11){\makebox(0,0)[bl]{\(O_{1}\)}}

\put(50.8,11){\makebox(0,0)[bl]{\(G_{1}\)}}

\end{picture}
\end{center}

\begin{center}
\begin{picture}(59,28)

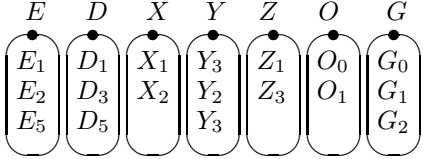
\put(03.5,00){\makebox(0,0)[bl] {Fig. 20. Superstructure of
solutions}}




\put(05,22){\circle*{1.5}} \put(13,22){\circle*{1.5}}
\put(21,22){\circle*{1.5}} \put(29,22){\circle*{1.5}}
\put(37,22){\circle*{1.5}} \put(45,22){\circle*{1.5}}
\put(53,22){\circle*{1.5}}

\put(04,24){\makebox(0,0)[bl]{\(E\)}}
\put(12,24){\makebox(0,0)[bl]{\(D\)}}
\put(20,24){\makebox(0,0)[bl]{\(X\)}}
\put(28,24){\makebox(0,0)[bl]{\(Y\)}}
\put(35,24){\makebox(0,0)[bl]{\(Z\)}}

\put(43,24){\makebox(0,0)[bl]{\(O\)}}
\put(52,24){\makebox(0,0)[bl]{\(G\)}}


\put(05,14){\oval(07,16)} \put(13,14){\oval(07,16)}
\put(21,14){\oval(07,16)} \put(29,14){\oval(07,16)}
\put(37,14){\oval(07,16)} \put(45,14){\oval(07,16)}
\put(53,14){\oval(07,16)}

\put(02.8,17){\makebox(0,0)[bl]{\(E_{1}\)}}
\put(02.8,13){\makebox(0,0)[bl]{\(E_{2}\)}}
\put(02.8,09){\makebox(0,0)[bl]{\(E_{5}\)}}

\put(10.8,17){\makebox(0,0)[bl]{\(D_{1}\)}}
\put(10.8,13){\makebox(0,0)[bl]{\(D_{3}\)}}
\put(10.8,09){\makebox(0,0)[bl]{\(D_{5}\)}}

\put(18.8,17){\makebox(0,0)[bl]{\(X_{1}\)}}
\put(18.8,13){\makebox(0,0)[bl]{\(X_{2}\)}}

\put(26.8,17){\makebox(0,0)[bl]{\(Y_{3}\)}}
\put(26.8,13){\makebox(0,0)[bl]{\(Y_{2}\)}}
\put(26.8,09){\makebox(0,0)[bl]{\(Y_{3}\)}}

\put(34.8,17){\makebox(0,0)[bl]{\(Z_{1}\)}}
\put(34.8,13){\makebox(0,0)[bl]{\(Z_{3}\)}}

\put(42.8,17){\makebox(0,0)[bl]{\(O_{0}\)}}
\put(42.8,13){\makebox(0,0)[bl]{\(O_{1}\)}}

\put(50.8,17){\makebox(0,0)[bl]{\(G_{0}\)}}
\put(50.8,13){\makebox(0,0)[bl]{\(G_{1}\)}}
\put(50.8,09){\makebox(0,0)[bl]{\(G_{2}\)}}


\end{picture}
\end{center}

 The extension procedure is the following.
 Table 16 contains addition operations and their estimates
 (scales \([1,3]\), expert judgment).

\begin{center}
\begin{picture}(51,35)

\put(04.5,31){\makebox(0,0)[bl] {Table 16. Addition operations}}

\put(0,0){\line(1,0){51}} \put(0,19){\line(1,0){51}}
\put(0,29){\line(1,0){51}}


\put(0,0){\line(0,1){29}} \put(05,0){\line(0,1){29}}
\put(21,0){\line(0,1){29}} \put(34,0){\line(0,1){29}}
\put(42,0){\line(0,1){29}} \put(51,0){\line(0,1){29}}

\put(2,25){\makebox(0,0)[bl]{\(i\)}}
\put(6,24.5){\makebox(0,0)[bl]{Operation}}

\put(22,24.5){\makebox(0,0)[bl]{Binary}}
\put(22,21){\makebox(0,0)[bl]{variable}}

\put(34.6,25){\makebox(0,0)[bl]{Cost}}
\put(37,21){\makebox(0,0)[bl]{\(a_{i}\)}}

\put(42.6,25){\makebox(0,0)[bl]{Profit}}
\put(45,21){\makebox(0,0)[bl]{\(c_{i}\)}}



\put(1.6,14){\makebox(0,0)[bl]{\(1\)}}

\put(6,13.5){\makebox(0,0)[bl]{\(E_{2} \Rightarrow E_{5}\)}}

\put(26,14){\makebox(0,0)[bl]{\(x_{1}\)}}

\put(1.6,10){\makebox(0,0)[bl]{\(2\)}}

\put(6,09.5){\makebox(0,0)[bl]{\(Y_{1} \Rightarrow Y_{3}\)}}

\put(26,10){\makebox(0,0)[bl]{\(x_{2}\)}}

\put(1.6,6){\makebox(0,0)[bl]{\(3\)}}

\put(6,5.5){\makebox(0,0)[bl]{\(Z_{1} \Rightarrow Z_{3}\)
 }}

\put(26,6){\makebox(0,0)[bl]{\(x_{2}\)}}

\put(1.6,2){\makebox(0,0)[bl]{\(4\)}}

\put(6,01.5){\makebox(0,0)[bl]{\(G_{1} \Rightarrow G_{2}\)}}

\put(26,02){\makebox(0,0)[bl]{\(x_{4}\)}}


\put(37,14){\makebox(0,0)[bl]{\(3\)}}

\put(46,14){\makebox(0,0)[bl]{\(3\)}}


\put(37,10){\makebox(0,0)[bl]{\(1\)}}

\put(46,10){\makebox(0,0)[bl]{\(3\)}}


\put(37,06){\makebox(0,0)[bl]{\(2\)}}

\put(46,06){\makebox(0,0)[bl]{\(1\)}}


\put(37,02){\makebox(0,0)[bl]{\(2\)}}

\put(46,02){\makebox(0,0)[bl]{\(3\)}}

\end{picture}
\end{center}

  The addition problem (simplified knapsack problem) is:
 \[\max \sum_{i=1}^{4} c_{i} x_{i} ~~~
%
  s.t. \sum_{i=1}^{4}  ~a_{i} x_{i} \leq b, ~~x_{i} \in \{0,1\}.\]
%
%
 Examples of the obtained resultant aggregated solutions  are
  (a simple greedy algorithm was used;
  the algorithm is based on ordering of elements by
  \(c_{i}/a_{i}\)):

 (1) \(b_{1}=5\):~
 ~(\(x_{1} = 0\), \(x_{2} = 1\), \(x_{3} = 1\), \(x_{4} = 1\)),

 \(S'_{b_{1}} =
 E_{2}\star D_{1}\star X_{1}\star Y_{3}\star Z_{3}
 \star O_{1}\star G_{1} \);

 (2) \(b_{2}=6\):~
 ~(\(x_{1} = 1\), \(x_{2} = 1\), \(x_{3} = 0\), \(x_{4} = 1\)),

 \(S'_{b_{2}} =
 E_{5}\star D_{1}\star X_{1}\star Y_{3}\star Z_{1}
 \star O_{1}\star G_{2} \).

%

 The procedure of new design  is the following.
 Table 17 contains design alternatives
 and their estimates
 (scales \([1,5]\), expert judgment).
 The design alternatives correspond
 to superstructure (Fig. 20).

\begin{center}
\begin{picture}(51,92)

\put(05,88){\makebox(0,0)[bl] {Table 17. Design alternatives}}

\put(0,0){\line(1,0){51}} \put(0,75){\line(1,0){51}}
\put(0,85){\line(1,0){51}}


\put(0,0){\line(0,1){85}} \put(05,0){\line(0,1){85}}
\put(21,0){\line(0,1){85}} \put(34,0){\line(0,1){85}}
\put(42,0){\line(0,1){85}} \put(51,0){\line(0,1){85}}

\put(2,81){\makebox(0,0)[bl]{\(\kappa\)}}

\put(5.6,80.5){\makebox(0,0)[bl]{Design}}
\put(5.6,77){\makebox(0,0)[bl]{alternative}}

\put(22,80.5){\makebox(0,0)[bl]{Binary}}
\put(22,77){\makebox(0,0)[bl]{variable}}

\put(34.6,81){\makebox(0,0)[bl]{Cost}}
\put(36,76.5){\makebox(0,0)[bl]{\(a_{ij}\)}}

\put(42.6,81){\makebox(0,0)[bl]{Profit}}
\put(44.5,76.5){\makebox(0,0)[bl]{\(c_{ij}\)}}


\put(1.6,70){\makebox(0,0)[bl]{\(1\)}}

\put(11,69.5){\makebox(0,0)[bl]{\(E_{1} \)}}

\put(25,70){\makebox(0,0)[bl]{\(x_{11}\)}}


\put(1.6,66){\makebox(0,0)[bl]{\(2\)}}

\put(11,65.5){\makebox(0,0)[bl]{\(E_{2} \)}}

\put(25,66){\makebox(0,0)[bl]{\(x_{12}\)}}


\put(1.6,62){\makebox(0,0)[bl]{\(3\)}}

\put(11,61.5){\makebox(0,0)[bl]{\(E_{5}\)}}

\put(25,62){\makebox(0,0)[bl]{\(x_{13}\)}}


\put(1.6,58){\makebox(0,0)[bl]{\(4\)}}

\put(11,57.5){\makebox(0,0)[bl]{\(D_{1}\) }}

\put(25,58){\makebox(0,0)[bl]{\(x_{21}\)}}


\put(1.6,54){\makebox(0,0)[bl]{\(5\)}}

\put(11,53.5){\makebox(0,0)[bl]{\(D_{3} \)}}

\put(25,54){\makebox(0,0)[bl]{\(x_{22}\)}}


\put(1.6,50){\makebox(0,0)[bl]{\(6\)}}

\put(11,49.5){\makebox(0,0)[bl]{\(D_{5} \)}}

\put(25,50){\makebox(0,0)[bl]{\(x_{23}\)}}


\put(1.6,46){\makebox(0,0)[bl]{\(7\)}}

\put(11,45.5){\makebox(0,0)[bl]{\(X_{1}\)}}

\put(25,46){\makebox(0,0)[bl]{\(x_{31}\)}}


\put(1.6,42){\makebox(0,0)[bl]{\(8\)}}

\put(11,41.5){\makebox(0,0)[bl]{\(X_{2}\) }}

\put(25,42){\makebox(0,0)[bl]{\(x_{32}\)}}


\put(1.6,38){\makebox(0,0)[bl]{\(9\)}}

\put(11,37.5){\makebox(0,0)[bl]{\(Y_{1} \)}}

\put(25,38){\makebox(0,0)[bl]{\(x_{41}\)}}


\put(0.6,34){\makebox(0,0)[bl]{\(10\)}}

\put(11,33.5){\makebox(0,0)[bl]{\(Y_{2} \)}}

\put(25,34){\makebox(0,0)[bl]{\(x_{42}\)}}


\put(0.6,30){\makebox(0,0)[bl]{\(11\)}}

\put(11,29.5){\makebox(0,0)[bl]{\(Y_{3}\)}}

\put(25,30){\makebox(0,0)[bl]{\(x_{43}\)}}


\put(0.6,26){\makebox(0,0)[bl]{\(12\)}}

\put(11,25.5){\makebox(0,0)[bl]{\(Z_{1}\) }}

\put(25,26){\makebox(0,0)[bl]{\(x_{51}\)}}


\put(0.6,22){\makebox(0,0)[bl]{\(13\)}}

\put(11,21.5){\makebox(0,0)[bl]{\(Z_{3} \)}}

\put(25,022){\makebox(0,0)[bl]{\(x_{52}\)}}


\put(0.6,18){\makebox(0,0)[bl]{\(14\)}}

\put(11,17.5){\makebox(0,0)[bl]{\(O_{0} \)}}

\put(25,18){\makebox(0,0)[bl]{\(x_{61}\)}}


\put(0.6,14){\makebox(0,0)[bl]{\(15\)}}

\put(11,013.5){\makebox(0,0)[bl]{\(O_{1}\)}}

\put(25,14){\makebox(0,0)[bl]{\(x_{62}\)}}


\put(0.6,10){\makebox(0,0)[bl]{\(16\)}}

\put(11,9.5){\makebox(0,0)[bl]{\(G_{0}\) }}

\put(25,10){\makebox(0,0)[bl]{\(x_{71}\)}}


\put(0.6,6){\makebox(0,0)[bl]{\(17\)}}

\put(11,05.5){\makebox(0,0)[bl]{\(G_{1} \)}}

\put(25,06){\makebox(0,0)[bl]{\(x_{72}\)}}


\put(0.6,2){\makebox(0,0)[bl]{\(18\)}}

\put(11,01.5){\makebox(0,0)[bl]{\(G_{2} \)}}

\put(25,02){\makebox(0,0)[bl]{\(x_{73}\)}}



\put(37,70){\makebox(0,0)[bl]{\(3\)}}
\put(46,70){\makebox(0,0)[bl]{\(3\)}}


\put(37,66){\makebox(0,0)[bl]{\(3\)}}
\put(46,66){\makebox(0,0)[bl]{\(4\)}}


\put(37,62){\makebox(0,0)[bl]{\(4\)}}
\put(46,62){\makebox(0,0)[bl]{\(5\)}}


\put(37,58){\makebox(0,0)[bl]{\(2\)}}
\put(46,58){\makebox(0,0)[bl]{\(2\)}}


\put(37,54){\makebox(0,0)[bl]{\(3\)}}
\put(46,54){\makebox(0,0)[bl]{\(3\)}}


\put(37,50){\makebox(0,0)[bl]{\(5\)}}
\put(46,50){\makebox(0,0)[bl]{\(4\)}}


\put(37,46){\makebox(0,0)[bl]{\(3\)}}
\put(46,46){\makebox(0,0)[bl]{\(4\)}}


\put(37,42){\makebox(0,0)[bl]{\(2\)}}
\put(46,42){\makebox(0,0)[bl]{\(3\)}}


\put(37,38){\makebox(0,0)[bl]{\(2\)}}
\put(46,38){\makebox(0,0)[bl]{\(2\)}}


\put(37,34){\makebox(0,0)[bl]{\(2\)}}
\put(46,34){\makebox(0,0)[bl]{\(3\)}}


\put(37,30){\makebox(0,0)[bl]{\(3\)}}
\put(46,30){\makebox(0,0)[bl]{\(4\)}}


\put(37,26){\makebox(0,0)[bl]{\(1\)}}
\put(46,26){\makebox(0,0)[bl]{\(1\)}}


\put(37,22){\makebox(0,0)[bl]{\(2\)}}
\put(46,22){\makebox(0,0)[bl]{\(2\)}}


\put(37,18){\makebox(0,0)[bl]{\(1\)}}
\put(46,18){\makebox(0,0)[bl]{\(1\)}}


\put(37,14){\makebox(0,0)[bl]{\(2\)}}
\put(46,14){\makebox(0,0)[bl]{\(3\)}}


\put(37,10){\makebox(0,0)[bl]{\(1\)}}
\put(46,10){\makebox(0,0)[bl]{\(1\)}}


\put(37,06){\makebox(0,0)[bl]{\(2\)}}
\put(46,06){\makebox(0,0)[bl]{\(3\)}}


\put(37,02){\makebox(0,0)[bl]{\(2\)}}
\put(46,02){\makebox(0,0)[bl]{\(4\)}}

\end{picture}
\end{center}

 It is assumed design alternatives for different
 product components are compatible.
 Thus, multiple choice problem for the new design is used:
 \[\max \sum_{i=1}^{7}  \sum_{j=1}^{q_{i}}   c_{ij} x_{ij}
 ~~~s.t.~ \sum_{i=1}^{7}  \sum_{j=1}^{q_{i}}   a_{ij} x_{ij} \leq
 b,
 \]
 \[\sum_{j=1}^{q_{i}}   x_{ij} = 1 ~~  \forall i=\overline{1,7},
  ~~x_{ij} \in \{0,1\}.\]
%
%
 Clearly, \(q_{1} = 3\), \(q_{2} = 3\), \(q_{3} = 2\), \(q_{4} = 3\),
 \(q_{5} = 2\), \(q_{6} = 2\), \(q_{7} = 3\).
 Examples of the obtained resultant aggregated solutions  are
  (a simple greedy algorithm was used;
 the algorithm is based on ordering of elements by
  \(c_{i}/a_{i}\)):

 (1) \(b^{1}=14\):~
 (\(x_{12} = 1\), \(x_{21} = 1\), \(x_{32} = 1\), \(x_{42} = 1\),
 \(x_{51} = 1\), \(x_{62} = 1\), \(x_{73} = 1\)),
 ~\(S''_{b^{1}} =
 E_{2}\star D_{1}\star X_{2}\star Y_{2}\star Z_{1}
 \star O_{1}\star G_{2} \);

 (2) \(b^{2}=17\):~
 (\(x_{13} = 1\), \(x_{22} = 1\), \(x_{31} = 1\), \(x_{41} = 1\),
 \(x_{52} = 1\), \(x_{62} = 1\), \(x_{73} = 1\)),
 ~\(S''_{b^{2}} =
 E_{5}\star D_{3}\star X_{1}\star Y_{3}\star Z_{3}
 \star O_{1}\star G_{2} \).

%


\section{Conclusion}

 In the paper,
 prospective frameworks for electronic shopping of modular
 products
 are suggested and examined
 (selection, composition/synthesis, aggregation).
 A special composite hierarchical structure
 for modular products
 is used:
 tree-like system model,
 design alternatives for product components,
 priorities of the design alternatives,
 estimates of compatibility between design alternatives.
 Solving procedures are based on combinatorial solving frameworks
 (multicriteria ranking, knapsack-like problems,
 hierarchical morphological design, aggregation).
%
 The suggested approaches
 have been illustrated by
 simplified applied realistic examples.

 In the future, it may be  reasonable to consider
 the following research directions:

 {\it 1.} investigation of other applications;

%
 {\it 2.} taking into account user's/customer's profiles;

%
 {\it 3.} usage of multicriteria knapsack problem
 and multicriteria multiple choice problem;

{\it 4.} examination of various kinds of proximity between
 composite products;

 {\it 5.} consideration of support tools to
 design  product structures;
 and
%
%
%
%
%

%
 {\it 6.} usage of fuzzy set approaches and AI techniques
  in the examined product design problems.
%
%



\end{document}